**Title:**

# Dielectric Metasurfaces for Complete and Independent Control of the Optical Amplitude and Phase

**Running Title:**

Phase-Amplitude Dielectric Metasurfaces

**Authors**:

*Adam C. Overvig*[1]*, Sajan Shrestha*[1]*, Stephanie C. Malek*[1]*, Ming Lu*[2]*, Aaron Stein*[2]*, Changxi Zheng*[3]*, and Nanfang Yu*[*1]

[1]Department of Applied Physics and Applied Mathematics, Columbia University, New York, NY 10027, USA
[2]Center for Functional Nanomaterials, Brookhaven National Laboratory, Upton, NY 11973, USA
[3]Department of Computer Science, Columbia University, New York, NY 10027, USA

**E-mails:**

Adam C. Overvig: aco2124@columbia.edu
Sajan Shrestha: ss4554@columbia.edu
Stephanie C. Malek: scm2185@columbia.edu
Ming Lu: mlu@bnl.gov
Aaron Stein: stein@bnl.gov
Changxi Zheng: cxz@cs.columbia.edu
Nanfang Yu: ny2214@columbia.edu

**Corresponding author:**

Nanfang Yu
201 S.W. Mudd
Mail Code 4701
Telephone: (212) 854-2196 / (212) 854-9940
Fax: (212) 854-8257

## Abstract

Metasurfaces are optically thin metamaterials that promise complete control of the wavefront of light but are primarily used to control only the phase of light. Here, we present an approach, simple in concept and in practice, that uses meta-atoms with a varying degree of form birefringence and rotation angles to create high-efficiency dielectric metasurfaces that control both the optical amplitude and phase at one or two frequencies. This opens up applications in computer-generated holography, allowing faithful reproduction of both the phase and amplitude of a target holographic scene without the iterative algorithms required in phase-only holography. We demonstrate all-dielectric metasurface holograms with independent and complete control of the amplitude and phase at up to two optical frequencies simultaneously to generate two- and three-dimensional holographic objects. We show that phase-amplitude metasurfaces enable a few features not attainable in phase-only holography; these include creating artifact-free two-dimensional holographic images, encoding phase and amplitude profiles separately at the object plane, encoding intensity profiles at the metasurface and object planes separately, and controlling the surface textures of three-dimensional holographic objects.

## Introduction

Structuring materials for arbitrary control of an optical wavefront is a long sought-after capability, enabling any physically possible linear optical functionality. Four key properties of a light wave are the amplitude, phase, polarization, and optical impedance. The ability to tune these properties at specific frequencies with subwavelength spatial resolution is the goal and promise of a class of metamaterials known as "metasurfaces", flat optical components composed of subwavelength structures with tailored optical responses[1]. By engineering these individual structures, or "meta-atoms", and properly arranging them on a surface, a wide range of desired linear optical functionalities can be achieved[2-5].

In practice, device functionality is limited by our ability to completely control these four properties arbitrarily and independently. This limitation comes down to the challenge of engineering the individual meta-atoms with widely varying desired responses at desired frequencies within a single achievable fabrication scheme. For this reason, most of the effort in the field of metasurfaces has focused on a single property at a time. Since phase is arguably the single most important property for wavefront control, metasurfaces engineering the phase profile of a wavefront dominate the published works[1-5]. While metallic scatterers are often used due to their strong light-matter interactions[6-10], to overcome the inherent optical losses involved with metals, lossless dielectric material platforms are commonly employed for high-efficiency phase control[11-19].

Expanding the gamut of achievable flat optical devices requires control of more than just the phase. For this reason, recent efforts have pushed for simultaneous control of more than one parameter at a time. A number of works have shown the flexibility of controlling the phase and polarization independently, enabling devices such as polarimeters[20], polarization-dependent lensing[13,21,22], and polarization-dependent holography[13,15,23,24]. Of considerable recent interest is controlling the phase at different frequencies independently, enabling multiwavelength or achromatic metasurfaces[25-29], dispersion-engineered devices[26], and multicolor holograms[30-34].

The most general linear optical device is the hologram, originally conceived as a microscopic principle encoding the amplitude and phase simultaneously[35,36]. Due to constraints in the ability to control an optical wavefront, metasurface holography is conventionally performed with a meta-atom library that controls only the phase[37]. Recent efforts have demonstrated meta-atom geometries allowing simultaneous amplitude and phase control and explored the benefits thereof for holography[38-41]. However, these efforts have been limited in efficiency or achieve results with unnecessary complexity.

Here, we present a metasurface platform with arbitrary and simultaneous control of the amplitude and phase at telecommunication frequencies in a transmission-type device. The amplitude is controlled by varying the conversion efficiency of circularly polarized light of one handedness into the circular polarization of the opposite handedness via structurally birefringent meta-atoms, while the phase is controlled by the in-plane orientation of the meta-atoms. This approach is a generalization of the well-studied metasurface platform employing the "geometric" or "Pancharatnam-Berry" phase, and we stress the conceptual and practical simplicity of this approach for achieving simultaneous and independent control of the amplitude and phase. This approach is easily generalizable to visible frequencies, and the fabrication of these dielectric metasurfaces is CMOS compatible. To demonstrate the advantage of simultaneous amplitude and phase control, we compare computer-generated holograms implemented with phase-and-amplitude (PA) metasurfaces and holograms implemented with phase-only (PO) metasurfaces and show that only the former are capable of creating artifact-free holographic images. To demonstrate the ability of PA holography to enable artistically interesting and complex scenes, we create metasurface holograms to generate high-fidelity three-dimensional (3D) holographic objects with distinct surface textures. To explore the utility of having two degrees of freedom per pixel, we create metasurfaces controlling both the amplitude and phase at the object plane and create a metasurface that has a grayscale image in the amplitude distribution and whose phase distribution produces a distinct holographic image at the object plane. Finally, we extend this simple scheme to include structural dispersion engineering of meta-atoms and demonstrate control of the phase and amplitude at two colors simultaneously.

**Results**

A long-employed approach for spatially varying the phase of light is to use the geometric phase[16,18,42], which is associated with the orientation of the linear polarization basis used to decompose circularly polarized light and can be simply altered by changing the orientation of the "fast axis" of a birefringent material. In the context of metasurfaces, "structural birefringence" is realized with metallic or dielectric scatterers with a different optical response in one in-plane direction compared to the orthogonal in-plane direction, and the orientation of these in-plane directions is tuned to control the phase of output circularly polarized light.

The operation of this metasurface on a wavefront is best described by using the Jones calculus[43]. In metasurfaces based on the geometric phase, the outgoing polarization state is modified from an incoming state as:

$$|\psi_2\rangle = \Gamma(-\alpha)M\Gamma(\alpha)|\psi_1\rangle, \qquad (1)$$

where $|\psi_1\rangle$ and $|\psi_2\rangle$ are Jones vectors in an $(x, y)$ basis describing the incoming and outgoing polarization states, respectively, $\Gamma(\alpha)$ is the $2 \times 2$ matrix rotating a unit vector in-plane by an angle $\alpha$, and $M$ is a matrix accounting for the outgoing amplitudes ($A_o$ and $A_e$) and phases ($\phi_o$ and $\phi_e$) for light polarized along the ordinary and extraordinary axes, respectively:

$$M = \begin{bmatrix} A_o e^{i\phi_o} & 0 \\ 0 & A_e e^{i\phi_e} \end{bmatrix}. \qquad (2)$$

Here, we consider the accumulated phase to be due to propagation within a meta-atom, which can be thought of as a short, vertically oriented dielectric waveguide, and assume unity transmittance (or forward scattering efficiency, $\eta_{forward}$) for both polarizations, which corresponds to $A_o = A_e = 1$. We can simplify $M$ and write the relevant phases in terms of the effective refractive indices $n_o$ and $n_e$, meta-atom height $d$, and free-space wavevector $k_0 = 2\pi/\lambda$ corresponding to wavelength $\lambda$:

$$\phi_{o,e} = k_0 n_{o,e} d. \qquad (3)$$

We take the incident polarization state to be circularly polarized light of one handedness (here, left circularly polarized, or LCP, with the Jones vector denoted as $|L\rangle$) and the signal (outgoing) state to be the opposite handedness (here, right circularly polarized, or RCP, with the Jones vector denoted as $|R\rangle$). As schematically depicted in **Figure 1a**, a polarization filter in the experimental setup selects only the RCP component of the outgoing wave, yielding a signal, $S$ (see Supporting Information Section S1 for a detailed derivation):

$$S = \langle R|\Gamma(-\alpha)M\Gamma(\alpha)|L\rangle = i \sin\left(\frac{k_0 d(n_o - n_e)}{2}\right) \exp\left(i\left(\frac{k_0 d(n_o + n_e)}{2} + 2\alpha\right)\right). \qquad (4)$$

This signal is therefore a complex value with both an amplitude and a phase. The amplitude is solely dependent on the sine term, the argument of which depends in particular on the degree of birefringence of the meta-atom, $(n_o - n_e)$. This amplitude can also be thought of as the conversion amplitude, that is,

$$\eta_{conversion} = \sin\left(\frac{k_0 d(n_o - n_e)}{2}\right), \qquad (5)$$

from LCP to RCP. It is unity when $|n_0 - n_e|d = \lambda/2$ and is zero when the meta-atom has no birefringence, that is, $|n_0 - n_e|d = 0$. Every other amplitude in between is achievable by varying the degree of birefringence between these two extremes.

The conventional choice for metasurfaces based on the geometric phase is to tune the birefringence to the half-wave-plate condition, yielding the maximum optical amplitude. Then, the optical phase is controlled through the rotation angle, $\alpha$. Here, we generalize this approach by creating a meta-atom library utilizing both $\alpha$ and the degree of birefringence of the meta-atoms, as visualized in **Figure 1b**. The amplitude is controlled entirely by the degree of form birefringence, while the phase is a sum of the propagation phase, $\frac{k_0 d(n_o+n_e)}{2}$, and the geometric phase $2\alpha$ (Equation 4). In this way, both the amplitude and phase can be completely and independently controlled.

The action this meta-atom library performs on input circularly polarized light can be visualized by paths along the Poincaré sphere (**Figure 1c**). The incident LCP light is placed at the south pole of the Poincaré sphere. The birefringence of the meta-atom determines the "latitude" of the output state, while the rotation angle $\alpha$ determines the "longitude" on the Poincaré sphere. In this way, incident LCP light can be converted into any polarization state (see Supporting Information Section S2). With the addition of a polarization filter (selecting for RCP light and absorbing the remaining LCP light), the output state on the Poincaré sphere is mapped to the amplitude and phase of the RCP light.

For a proof-of-concept implementation, we choose an operating wavelength of $\lambda = 1.55\ \mu m$ and a CMOS-compatible platform of amorphous silicon ($a$-Si) metasurfaces on fused silica substrates. The metasurface holograms consist of a square lattice of meta-atoms with rectangular in-plane cross-sections, with the geometric parameters defined in **Figure 1d**. A lattice constant of $P = 650\ nm$ and meta-atom height of $d = 800\ nm$ are chosen so that for a large variation of $W_x$ and $W_y$ (in-plane widths of the meta-atoms), the forward scattering amplitudes, $\eta_{forward}$, for both $x$ and $y$ polarized light are near unity (see Supporting Information Section S3). This ensures that $A_o \cong A_e \cong 1$ and that the conversion amplitude is identical to the amplitude of the output signal:

$$|S| = \eta_{forward}\eta_{conversion} \cong \sin\left(\frac{k_0 d(n_o-n_e)}{2}\right). \qquad (6)$$

To find suitable combinations of $W_x$ and $W_y$ of the target meta-atom library, finite-difference time-domain (FDTD, Lumerical Solutions) simulations are performed, and a contour through the simulated parameter space is chosen that closely satisfies the condition of $\eta_{forward} = 1$ while providing $\eta_{conversion}$ that continuously varies from 0 to 1. The specific chosen contour has $W_x = 200\ nm$ and $W_y$ varying from $200\ nm$ to $480\ nm$ (refer to Supporting Information Section S3).

The amplitude and phase of the RCP component of the output are then recorded for each combination of $W_y$ and $\alpha$, as shown in **Figure 1e**. Note that the converted amplitude is essentially independent of the orientation angle, indicating that the effect of coupling between neighboring meta-atoms on effective refractive indices $n_0$ and $n_e$ is negligible and validating the absence of $\alpha$ in Equation 6. For ease of use, the simulation results are inverted into a "look-up" table (**Figure 1f**) (see Supporting Information Section S4 for this process), wherein a desired amplitude and phase combination can be converted to the required geometric parameters, $W_y$ and $\alpha$. The successful inversion from **Figure 1e** to **Figure 1f** numerically demonstrates the arbitrary control of the amplitude and phase achieved by the meta-atom library.

To showcase the complete control of the amplitude and phase, computer-generated holograms (CGHs) are implemented experimentally. Five CGHs are demonstrated: the first generates a two-dimensional (2D) holographic image and demonstrates improved fidelity of the image produced with PA holography over those produced with two versions of PO holography (**Figure 2**); the second is a CGH that creates a simple 3D holographic scene consisting of a collection of points and demonstrates 3D holography by the dependence of the reconstructed holographic scene on the focal plane and observation angle of the imaging optics (**Figure 3**); the third CGH demonstrates the faithful reconstruction of a complex 3D holographic object (**Figure 4**); the fourth demonstrates the ability to separately encode the phase and amplitude at the object plane (**Figure 5**); and the fifth demonstrates the encoding of a holographic image with the phase distribution of a grayscale hologram, itself an image in the amplitude distribution (**Figure 6**). Detailed information about the CGHs can be found in Supporting Table S1.

To generate the 2D CGH, a target image (the Columbia Engineering logo) is discretized into dipole sources with amplitudes of 1 (corresponding to the area inside the logo) and 0 (corresponding to the background) and a uniform phase. A Gaussian filter is then applied to blur the sharp boundaries between the values of 0 and 1, as these boundaries represent information encoded at higher momenta than the free-space momentum (see Supporting Information Section S5 for the effect of skipping this blurring step). The interference of these dipole sources is recorded at a distance $D = 750\ \mu m$ from the target image, which corresponds to the location of the metasurface that will reconstruct this target image. The result is a complex transmission function, $\tilde{\tau}(x,y)$, required at the metasurface plane:

$$\tilde{\tau}(x,y) = \sum_{i,j} \frac{\exp(i\, k_0\, R_{ij}(x,y))}{R_{ij}(x,y)}, \qquad (7)$$

where $R_{ij}(x,y)$ is the distance from the $(i,j)th$ dipole source to a position $(x,y)$ on the metasurface. Finally, $\tilde{\tau}(x,y)$ is normalized: $\tilde{\tau}_{norm}(x,y) = \tilde{\tau}(x,y)/\,|\tilde{\tau}(x,y)|_{max}$. For the first PO hologram, the amplitudes are simply set to unity.

For the second PO hologram, which we refer to as the GS hologram, an alternate approach (called the Gerchberg-Saxton algorithm[44]) is used, which sets amplitude responses to unity and iteratively corrects the phase at the metasurface plane to generate the desired intensity distribution of the target image. No such iteration is necessary in the PA holography, as we can faithfully reproduce both the phase and amplitude of the desired hologram, the advantages and disadvantages of which are discussed below.

The resulting $\tilde{\tau}(x,y)$ for the PA, PO, and GS holograms are depicted in **Figure 2a-c.** The devices are fabricated using a CMOS-compatible process, described in Supporting Information Section S6. The resulting optical images of the 2D holograms are shown in **Figure 2d-f**. They consist of a layer of nanostructured amorphous silicon 0.8 μm in height patterned on a fused quartz substrate. The overall size of each hologram is 750 μm × 750 μm.

The reconstruction of each holographic image is performed both by numerical simulation (**Figure 2g-i**) and experimentally (**Figure 2j-l**, see Supporting Information Section S7 for experimental details). The improvement of the image quality in the PA compared to either PO or GS is readily apparent, reflecting the uncompromised reconstruction of a target image. The PO hologram can be seen to highlight the edges of the logo, suggesting that a role of amplitude variation in the PA hologram is to correctly modulate the amplitudes of the high spatial frequencies in the reconstructed image. This can be seen visually by comparing the $\tilde{\tau}(x,y)$ of PA and PO: where the outer edges of the hologram for the PA (representing a large bending angle) have low amplitude, the PO hologram must have unity amplitude. The GS hologram solves this limitation of the PO hologram by employing the iterative algorithm described above. However, it appears “grainy” or “splotchy” due to unwanted destructive interference within the logo boundaries, a well-known limitation of GS holography. The dependence on wavelength for a 2D PA and PO hologram is shown in **Figure S8**, demonstrating that the broad bandwidth of the geometric phase approach extends to PA holography.

A further showcase of the capabilities of PA holography can be seen in **Figure 3** and **Figure 4**, where 3D holography is demonstrated. **Figure 3a** shows $\tilde{\tau}(x, y)$ for generating a 3D coil, calculated by discretizing the coil into an array of dipole sources and recording their interference pattern at the metasurface plane. To show the depth of the 3D coil, three focal planes are chosen for experimental reconstruction, as depicted in **Figure 3b**. The individual dipole sources are discernible at the farthest focal plane of $300\ \mu m$, where the distribution of the dipoles is sparsest, while at the nearest focal plane of $100\ \mu m$, they are nearly continuous. As seen in **Figure 3c**, parallax is demonstrated by changing the viewing angle of the camera (maintaining normally incident light to the metasurfaces), with a recognizable image observed at an angle as high as 60° (approximate corresponding focal planes are drawn in **Figure 3c**). This verifies the true holographic nature of the experiment: the reconstruction simulates looking through a window into a virtual world populated by the 3D coil.

To demonstrate the ability of PA holography to enable more artistically interesting and complex scenes, a target 3D-modeled cow is converted into a hologram and then reconstructed. **Figure 4a** depicts the computation of $\tilde{\tau}(x, y)$ for generating the cow, computed with a simulation interfering light waves scattered off the 3D surface of the cow. This method of computer-generated holography, described in Supporting Information Section S9, includes realistic physical effects such as occlusion and surface textures. In particular, rough or smooth surface textures are simulated by choosing a random or uniform distribution of scattered phase over the surface of the cow. Three $\tilde{\tau}(x, y)$ are calculated in this manner and shown in **Figures 4b-d**. **Figure 4b** depicts $\tilde{\tau}(x, y)$ for a cow with a rough surface at an oblique perspective, while **Figures 4c,d** depict, respectively, $\tilde{\tau}(x, y)$ for a cow with a rough and a smooth surface from an edge-on perspective.

The optical reconstruction is performed both computationally (**Figure 4e**) and experimentally (**Figure 4f**). The excellent agreement, even in the details of the speckle pattern, affirms the fidelity with which the PA holography platform can capture effects such as surface roughness. See Supporting Information Section S10 for details on the simulated reconstruction. Reconstruction using an LED (linewidth ~120 nm centered around 1.55 μm) shows a reduction in the speckle contrast due to the increased bandwidth and incoherence of the source (see Supporting Information Section S11).

**Figures 4g,h** contain the simulated reconstructions of the rough and smooth cows, respectively, with the outline of the cow shown for reference. Notably, for the smooth cow, only the specular highlights (that is, the portions of the cow where the angle of incidence of the illumination is equal to the angle of observation) are apparent, while the rough cow shows a speckle pattern nearly filling the silhouette of the cow. We note that this speckle phenomenon is physically accurate and unintuitive only because of the rarity of coherent sources as the sole illumination source in everyday experience. The agreement with physical expectations demonstrates the control of PA holography over the surface texture of complex 3D holographic objects. Control over the surface texture is possible because of the simultaneous control of the object amplitude and phase, which is uniquely possible in PA holography.

PO holography uses only one degree of freedom (phase) at the hologram plane to control one degree of freedom (intensity) at the object plane. PA holography has no such limitations and, as seen in **Figure 5**, may separately encode the amplitude and phase of a holographic image. **Figures 5a,b** contain the complex transmission functions of two holograms that encode the same object intensity profiles but distinct object phase profiles (as shown in **Figures 5c,d**). Therefore, not only is the fidelity of the intensity profile improved in PA holography over PO holography (as seen in **Figure 2**) but also an entirely parallel channel of information (phase) can be faithfully encoded simultaneously. In this case, the phase profiles chosen are simple gradients, meaning that the holographic objects are observable from distinct angles. This is experimentally verified in **Figures 5e-h**, where the holographic images are formed only if the information projected by the holograms is within the range of angles collected by the imaging objective.

Another use of the two degrees of freedom present in PA holography is to control the amplitude profiles at two separate planes rather than the amplitude and phase at a single plane. To demonstrate this, we modify the GS algorithm to enforce a grayscale amplitude distribution (instead of the conventional uniform amplitude distribution) and iteratively recover the phase required to produce a target holographic image at the object plane given the chosen nonuniform amplitude distribution. In other words, as depicted in **Figure 6a**, the metasurface can be encoded with a grayscale image (**Figure 6b**) while simultaneously producing a holographic image (**Figure 6f**). The intensity and phase profiles of the resulting metasurface are shown in **Figures 6c,g**. The experimental reconstructions (**Figures 6e,i**) are in good agreement with the simulated reconstructions (**Figures 6d,h**), showing recognizable target images with artifacts inherent to GS holography (destructive interference due to a lack of phase control at each plane). Supporting Video S1 shows the transformation between the reconstructed images as the focal plane of the imaging setup is adjusted between the hologram and the object planes. Supporting Information Section S14 explores the trade-offs in image quality at the two planes and the qualitatively different nature of the "speckle" at the metasurface plane (born of the phase discontinuities) compared to that at the object plane (born of the rapidly changing phase profile).

Finally, we extend this simple approach to control the amplitude and phase independently at two separate wavelengths[34]. This represents control of four wavefront parameters simultaneously at each meta-atom and therefore requires more degrees of freedom in the meta-atom design than the two degrees of freedom (aspect ratio and orientation of rectangular meta-atoms) used above. We have shown previously that structural dispersion engineering of meta-atoms by widely varying their cross-sectional shapes (while retaining rotational symmetry or four-fold symmetry) can yield a library controlling the phase of a wide range of wavelengths at a time[29]. We extend this past effort to include form birefringence in the design of meta-atoms, allowing expansive control of the phase response of the ordinary and extraordinary polarizations at two wavelengths.

Specifically, four archetypes of meta-atoms supporting form birefringence are used, each representing a subclass of meta-atoms with the geometric degrees of freedom indicated by the arrows in **Figure 7a**. In addition, we (1) increase the thickness of the amorphous silicon layer from 0.8 μm to 1 μm to increase the range of phase dispersion resulting from propagation, (2) choose relatively widely separated wavelengths representing "red" ($\lambda$=1.65 μm) and "blue" ($\lambda$=0.94 μm) channels to enhance the dispersion of the optical response, and (3) set the input handedness of circularly polarized light in the "red" to be opposite that in the "blue" so that the dependence of the phase on $\alpha$ is opposite for each color (further expanding the range of responses possible).

The phase, $\phi_R$, and dispersion, $\phi_B - \phi_R$, due to propagation through the meta-atom library are depicted in **Figure 7a**, demonstrating dense and degenerate coverage of this space. This degeneracy (many meta-atoms providing the same phase dispersion but different amplitudes) is key, as the amplitude must also vary widely and independently. The geometric phase is an additional degeneracy in the phase to be exploited and can be included by analytical extension of the numerical simulations. To visually explore how well the combinations of amplitude and phase $(A_R, A_B, \phi_R, \phi_B)$ at the two wavelengths are achieved, **Figure 7b** breaks the amplitudes into bins of $(A_R, A_B)$ and plots the $(\phi_R, \phi_B)$ within each bin. The apparent filling of every space in the $(\phi_R, \phi_B)$ plot for every bin indicates that our meta-atom library can achieve every combination of $(A_R, A_B, \phi_R, \phi_B)$ up to the precision of the bins chosen. These high-aspect-ratio meta-atoms with widely varying cross-sections therefore provide four independent degrees of wavefront control within a monolithic fabrication scheme.

For a proof-of-concept demonstration, a target two-color image (**Figure 7g**) is converted as before into the required amplitude and phase on the metasurface plane at each wavelength (where the red channel of the image is used for $\lambda = 1.65\ \mu m$ and the blue channel of the image is used for $\lambda = 0.94\ \mu m$), as depicted in **Figure 7c** and **Figure 7d**, respectively. Example scanning electron micrographs of the fabricated devices are shown in **Figures 7e,f**, exemplifying the diversity of cross-sections optically encoding four independent variables at each pixel. The two-color experimental reconstruction (**Figure 7h**) is acquired by aligning the results with LCP excitation at $\lambda = 1.65\ \mu m$ (**Figure 7i**) and RCP excitation at $\lambda = 0.94\ \mu m$ (**Figure 7j**). We note that for the "red" wavelength there is a good agreement with the target image, while the "blue" wavelength shows significant, yet poorer agreement. We attribute the difference in performance across wavelengths primarily to the poorer accuracy of the assumptions for the smaller wavelength involved in producing the meta-atom library seen in **Figure 7b**. In particular, at the smaller wavelength, the structures support higher-order modes and resonances arising from the complex interactions thereof, which degrades the reliability of the "single-pass approximation"[45]. Due to the number of meta-atoms that need to be simulated (**Figure 7a** represents ~60,000 meta-atoms), more accurate characterizations of the response of each meta-atom represents a daunting computational problem. We therefore restrict ourselves to the present imperfect but computationally tractable solution.

**Discussion**

The advantages of PA over PO holographic metasurfaces are clear in the above demonstrations but merit a more detailed discussion. Notably, PO holography has the advantage of improved power efficiency. This comes from the fact that all of the light incident on the PO hologram contributes to the final image, unlike in PA holography, where the amplitude is continuously modulated between 0 and 1, filtering a portion of the power out. We note, however, that this reduction in efficiency is (1) highly case dependent (e.g., different illumination patterns and target holographic objects will use the input power differently) and (2) ambiguous in direct comparison to PO holography. In particular, there is a trade-off between the degree to which "ringing artifacts" can be suppressed (see Supporting Information Section S5) and the amount of power contributing to the final image: ringing artifacts (related to Gibb's overshoot) can be reduced at the cost of lower overall efficiency (see Supporting Information Section S12). The choice of what counts as sufficient elimination of the artifacts will therefore determine the maximum efficiency of the hologram, meaning that there is no unambiguous comparison between PO and PA holography, as PO holography involves no such choice. Indeed, PO holography can be thought of as the choice within PA holography with maximal efficiency at the cost of maximal artifacts.

The cost of the increased power efficiency in PO holography is at least threefold. First, a substantially lower density of information is encoded by a PO hologram compared to that by its PA counterpart. This is because a PO hologram controls only the phase at each pixel in the metasurface plane, while a PA hologram controls both the amplitude and phase, which has the consequence that the phase at the object plane can be independently controlled by a PA hologram (**Figure 5**) but not by a PO hologram. This could allow, for example, increasing the difficulty of counterfeiting in security applications by using holographic images of identical appearance (intensity) but with detectable differences in phase profile that require special equipment to decode, such as an interference-based apparatus. Furthermore, in an application involving holographic data storage, there is a multiplicative effect on the storable bits per pixel: a system capable of reading out $M$ distinct values of the phase from a PO hologram would allow the storage of $M$ states per pixel, while a system using a PA hologram that simultaneously reads out $N$ values of the amplitude would allow the storage of $M \times N$ states per pixel.

Second, although the phase is not recorded directly by a camera or the human eye, the phase distribution on the optical wavefront contributes to the visual textures of a virtual object. As an example, a diffuse surface will have a random phase, while a glossy surface has some degree of phase uniformity. This texture detail is lost (or must be mimicked) by the PO approach but effortlessly retained in the PA approach (**Figure 4**), where both the desired phase and amplitude of the holographic object are faithfully reproduced.

Third, a Gerchberg-Saxton-like algorithm is necessary to reduce the unwanted distortions to the image (seen in **Figure 2**). While straightforward for reconstructing simple 2D scenes, the computational requirements make general PO holography (such as reconstructing 2D and 3D scenes[46-48] with controlled textures) difficult and often impractical to implement, especially in dynamic holography. As shown in **Figures 3-4**, no correction algorithm is necessary in 3D PA holography, which retains complete phase and amplitude information in the final 3D holographic scene. In other words, PA holography is faithful to the original imagination of holography: the PA hologram generates the wavefront produced by a virtual object and therefore is effectively a window into a virtual world.

In conclusion, we have demonstrated metasurface holograms using low-loss dielectric metasurfaces operating in transmission mode with complete and independent phase and amplitude control at one and two wavelengths. Structural dispersion engineering of meta-atoms and the geometric phase are employed to enable control of up to four wavefront parameters at each pixel of the metasurface holograms. This design principle is a simple but powerful extension of the long-employed geometric-phase metasurfaces, opening up a degree of control over optical wavefronts useful in many applications. We implemented monochromatic 2D and 3D phase-amplitude holograms using a library of meta-atoms with rectangular cross-sections supporting a wide range of form birefringence. We showed that the quality of 2D phase-amplitude holographic images was significantly improved over that of phase-only holography. We also showed that a PA metasurface may encode entirely separate profiles of the phase and amplitude at the object plane and that, for 3D holographic objects, this allows surface textures to be straightforwardly realized. We demonstrated holography using a generalized GS algorithm enabling holographic encoding with a grayscale hologram. We further implemented 2D holograms providing complete control of the optical phase and amplitude at two colors simultaneously using a library of meta-atoms with complex cross-sectional shapes. This work offers a robust and generalizable method towards realizing the primary promise of metasurfaces: to manipulate an optical wavefront at will.

## Materials and Methods

The holograms are numerically generated by computing the interference of complex amplitude point sources composing the target object at a plane to be occupied by the metasurface. As detailed in Supporting Information S9, the complex 3D object is computed using Monte Carlo integration over the mesh of the cow, with the addition of a scattering phase to simulate surface textures.

As detailed in Section S10, the simulated reconstruction of holograms is performed using the convolution method in the Fourier domain using a propagation kernel of a point source and the complex transmission function of the metasurface.

Full-wave simulations of individual meta-atoms are carried out using commercial finite-difference time-domain (FDTD) software, Lumerical Solutions.

As detailed in Section S7, optical characterization is carried out by illuminating the holograms with either a laser diode or light-emitting diode of the proper wavelength. Light is then circularly polarized by a linear polarization combined with a quarter-wave plate (Thorlabs) and passed to the metasurface. The light is collected by a 10× or 100× near-infrared objective (Mitotoyu), passed through a polarization filter (Thorlabs), and directed towards a near-infrared camera (Princeton Instruments).

Fabrication is carried out at Brookhaven National Laboratory using standard planar fabrication technologies, detailed in Section S6. Chemical vapor deposition is used to grow 800 nm to 1000 nm of amorphous silicon on a quartz wafer. A double-layer of poly(methyl-methacrylate) is spun and baked at 180°C to serve as an electron-beam resist in a lift-off procedure. Electron beam lithography (JEOL) is carried out at 100 keV and 500 pA, with a base dose of 740 $\mu C/cm^2$ and appropriate proximity effect corrections (BEAMER). A mixture of 3:1 isopropyl alcohol to deionized water develops the exposed resist. A thin layer of alumina is deposited using electron-beam deposition, and the excess resist is stripped using a bath of N-Methyl-2-pyrrolidone (NMP) at 85°C for 4 hours. Finally, the pattern is transferred into the silicon layer by reactive ion etching.

**Author contributions**

A.C.O. conceived the concepts and executed the design work and numerical simulations. A.C.O. and S.S. carried out the fabrication and optical characterization. A.C.O., S.C.M., and C.Z. numerically generated the holograms. M.L., A.S., and N.Y. supervised the fabrication. N.Y. supervised and directed the research.

**Competing interests**

None

**Materials & Correspondence**

Correspondence should be directed to Nanfang Yu, E-mail: ny2214@columbia.edu

**Acknowledgements**

The work was supported by the Defense Advanced Research Projects Agency (Young Faculty Award grant no. D15AP00111, Seedling grant no. HR0011-17-2-0017), the National Science Foundation (grant no. ECCS-1307948), and the Air Force Office of Scientific Research (grant no. FA9550-14-1-0389, through a Multidisciplinary University Research Initiative program). A.C.O. acknowledges support from the NSF IGERT program (grant no. DGE-1069240). Research was carried out in part at the Center for Functional Nanomaterials, Brookhaven

National Laboratory, which is supported by the US Department of Energy, Office of Basic Energy Sciences (contract no. DE-SC0012704).

## References

[1] Yu, N., Genevet, P., Kats, M. A., Aieta, F., Tetienne, J. P., Capasso, F. & Z. Gaburro, *Science* **334**, 333–337 (2011).

[2] Yu, N. & Capasso, F. *Nature Materials* **13**, 139−150 (2014).

[3] Kildishev, A. V., Boltasseva, A. & Shalaev, V. M. *Science* **339**, 1232009 (2013).

[4] Chen, H.-T., Taylor, A. J. & Yu, N. *Reports on Progress in Physics* **79**, 076401 (2016).

[5] He, Q., Sun, S., Xiao, S. & Zhou, L. *Advanced Optical Materials* **6**, 1800415 (2018).

[6] Sun, S., He, Q., Xiao, S., Xu, Q., Li, X. & Zhou, L. *Nature Materials* **11**, 426–431 (2012).

[7] Sun, S., Yang, K., Wang, C., Juan, T., Chen, W., Liao, C., He, Q., Xiao, S., Kung, W., Guo, G., Zhou, L. & Tsai, D. P. *Nano Letters* **12**, 6223–6229 (2012).

[8] Luo, W., Xiao, S., He, Q., Sun, S. & Zhou, L. *Advanced Optical Materials* **3**, 1102–1108 (2015).

[9] Meinzer, N., Barnes, W. L. & Hooper, I. R. *Nature Photonics* **8**, 889−898 (2014).

[10] Schuller, J. A., Barnard, E. S., Cai, W., Jun, Y. C., White, J. S. & Brongersma, M. L. *Nature Materials* **9**, 193–204 (2010).

[11] Shalaev, M. I., Sun, J., Tsukernik, A., Pandey, A., Nikolskiy, K. & Litchinitser, N. M. *Nano Letters* **15**, 6261−6266 (2015).

[12] Chong, K. E., Wang, L., Staude, I., James, A. R., Dominguez, J., Liu, S., Subramania, G. S., Decker, M., Neshev, D. N., Brener, I. & Kivshar, Y. S. *ACS Photonics* **3**, 514−519 (2016).

[13] Arbabi, A., Horie, Y., Bagheri, M. & Faraon, A. *Nature Nanotechnology* **10**, 937−944 (2015).

[14] Zhao, W., Jiang, H., Liu, B., Song, J., Jiang, Y., Tang, C. & Li, J. *Scientific Reports* **6**, 30613 (2016).

[15] Khorasaninejad, M., Ambrosio, A., Kanhaiya, P. & Capasso, F. *Science Advances* **2**, 1501258 (2016).

[16] Bomzon, Z., Kleiner, V. & Hasman, E. *Optics Letters* **26**, 1424−1426 (2001).

[17] Lalanne, P., Astilean, S., Chavel, P., Cambril, E. & Launois, H. *Optics Letters* **23**, 1081−1083 (1998).

[18] Bomzon, Z., Biener, G., Kleiner, V. & Hasman, E. *Optics Letters* **27**, 1141−1143 (2002).

[19] Decker, M., Staude, I., Falkner, M., Dominguez, J., Neshev, D. N., Brener, I., Pertsch, T. & Kivshar, Y. S. *Advanced Optical Materials* **3**, 813−820 (2015).

[20] Mueller, J. B., Leosson, K. & Capasso, F. *Optica* **3**, 42−47 (2016).

[21] Hasman, E., Kleiner, V., Biener, G. & Niv, A. *Applied Physics Letters* **82**, 328−330 (2003).

[22] Khorasaninejad, M., Chen, W. T., Zhu, A. Y., Oh, J., Devlin, R. C., Rousso, D. & Capasso, F. *Nano Letters* **16**, 4595−4600 (2016).

[23] Mueller, J. B., Rubin, N. A., Devlin, R. C., Groever, B. & Capasso, F. *Physical Review Letters* **118**, 113901 (2017).

[24] Wen, D., Yue, F., Li, G., Zheng, G., Chan, K., Chen, S., Chen, M., Li, K. F., Wong, P. W. H., Cheah, K. W., Pun, E. Y. B., Zhang, S. & Chen, X. *Nature Communications* **6**, 8241 (2015).

[25] Shrestha, S., Overvig, A. C. & Yu, N. "Broadband achromatic metasurface lenses," paper FM1H.3 in *Conference on Lasers and Electro-Optics*, San Jose, CA, 2017.

[26] Arbabi, E., Arbabi, A., Kamali, S. M., Horie, Y. & Faraon, A. *Optica* **4**, 625–632 (2017).

[27] Khorasaninejad, M., Shi, Z., Zhu, A. Y., Chen, W. T., Sanjeev, V., Zaidi A. & Capasso, F. *Nano Letters* **17**, 1819–1824 (2017).

[28] Wang, S., Wu, P. C., Su, V. C., Lai, Y. C., Chu, C. H., Lu, S. H., Chen, J., Xu, B., Kuan, C. H., Li, T., Zhu, S. & Tsai, D. P. *Nature Communications* **8**, 187 (2017).

[29] Shrestha, S., Overvig, A. C., Lu, M., Stein, A. & Yu, N. *Light: Science & Applications* **7**, 85 (2018).

[30] Li, X., Chen, L., Li, Y., Zhang, X., Pu, M., Zhao, Z., Ma, X., Wang, Y., Hong, M. & Luo, X. *Science Advances* **2**, 1601102 (2016).

[31] Wang, B., Dong, F., Li, Q. T., Yang, D., Sun, C., Chen, J., Song, Z., Xu, L., Chu, W., Xiao, Y. F., Gong, Q. & Li, Y. *Nano Letters* **16**, 5235–5240 (2016).

[32] Zhao, W., Liu, B., Jiang, H., Song, J., Pei, Y. & Jiang, Y. *Optics Letters* **41**, 147–150 (2015).

[33] Zhao, W., Jiang, H., Liu, B., Song, J., Jiang, Y., Tang, C. & Li, J. *Scientific Reports* **6**, 30613 (2016).

[34] Overvig, A., Shrestha, S., Xiao, C., Zheng, C. & Yu, N. "Two-color and 3D phase-amplitude modulation holograms," paper FF1F.6 in *Conference on Lasers and Electro-Optics*, San Jose, CA, 2018.

[35] Gabor, D. *Nature* **161**, 777–778 (1948).

[36] Gabor, D. *Vacuum* **16**, 313 (1966).

[37] Genevet, P. & Capasso, F. *Reports on Progress in Physics* **78**, 024401 (2015).

[38] Wang, Q., Zhang, X., Xu, Y., Gu, J., Li, Y., Tian, Z., Singh, R., Zhang, S., Han, J. & Zhang, W. *Scientific Reports* **6**, 32867 (2016).

[39] Overvig, A., Shrestha, S. Zheng, C. & Yu, N. "High-efficiency amplitude-phase modulation holograms based on dielectric metasurfaces," paper FM1H.4 in *Conference on Lasers and Opto-Electronics*, San Jose, CA, 2017.

[40] Lee, G. Y., Yoon, G., Lee, S. Y., Yun, H., Cho, J., Lee, K., Kim, H., Rho, J. & Lee, B. *Nanoscale* **10**, 4237–4245 (2018).

[41] Jia, S. L., Wan, X., Su, P., Zhao, Y. J. & Cui, T. J. *AIP Advances* **6**, 045024 (2016).

[42] Pancharatnam, S. *Proceedings of the Indian Academy of Science A* **44**, 247–262 (1956).

[43] Jones, R. C. *Journal of the Optical Society of America* **31**, 488–493 (1941).

[44] Gerchberg, R. W. & Saxton, W. O. *Optik* **35**, 237–246 (1972).

[45] Yang, J., Sell, D. & Fan, J. A. *Ann. Phys.* **530,** 1700302 (2018).

[46] Shabtay, G. *Optics Communications* **226**, 33–37 (2003).

[47] Whyte, G. & Courtial, J. *New Journal of Physics* **7**, 117 (2005).

[48] Xia, X. & Xia, J, *Chinese Physics B* **25**, 094204 (2016).

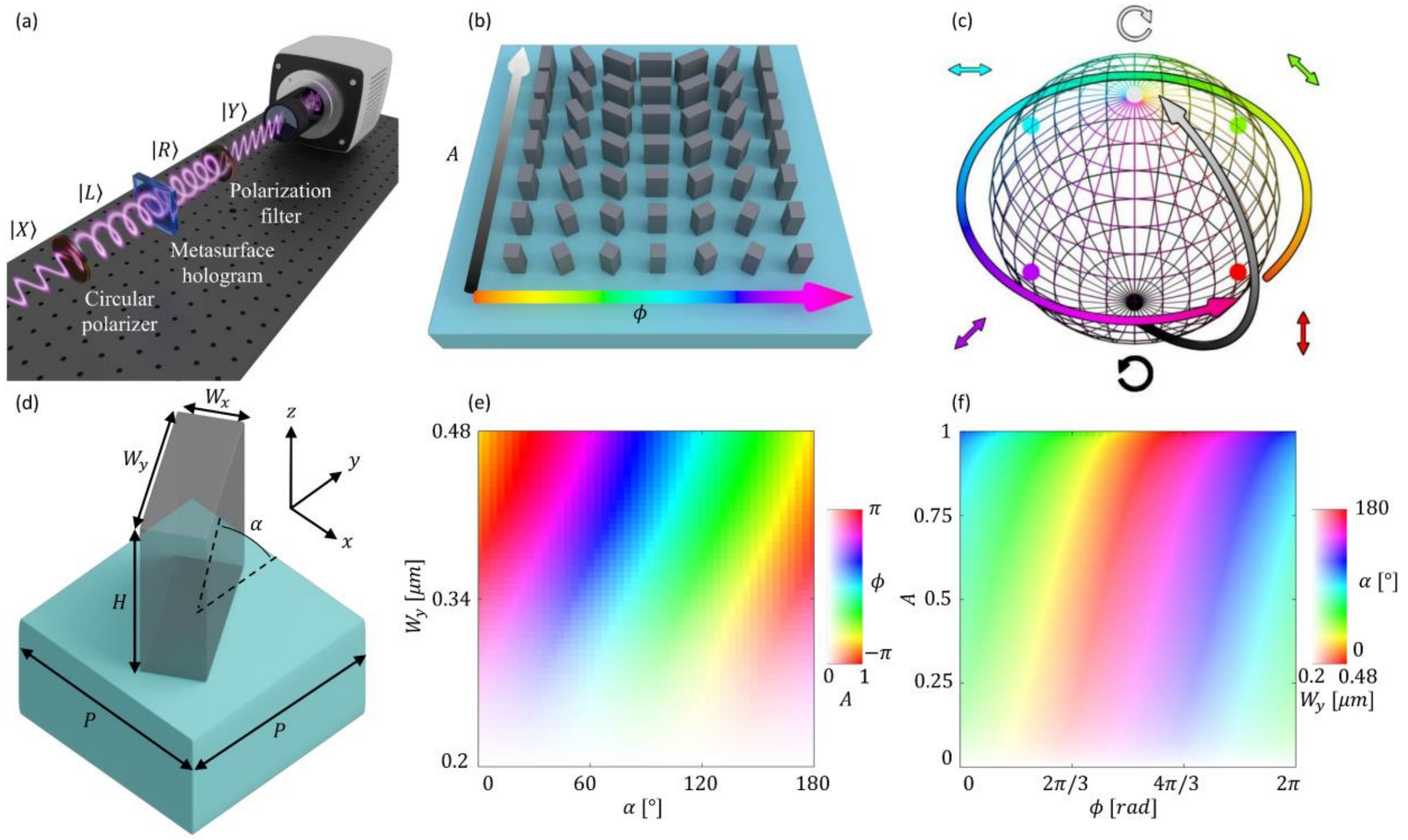

**Figure 1**. Two degrees of freedom enable independent and complete control of the optical amplitude and phase. (a) Schematic of the holographic experiment: circularly polarized light is partially converted by the metasurface to its opposite handedness and is then filtered by an analyzing polarization filter before forming an image on the camera. (b) Geometrical parameters of the meta-units sweep the amplitude (black-white gradient axis) and phase (rainbow axis) of the signal component of the output. (c) The unit cells in (b) can take incident left circularly polarized light (south pole) to any other point on the Poincaré sphere with near-unity efficiency representing two independent degrees of freedom controlled by the metasurface. (d) Geometric parameters of a meta-unit. (e) Full-wave simulations varying $W_y$ and $\alpha$ for $H = 800\ nm$, $W_x = 200\ nm$, $P = 650\ nm$, and $\lambda = 1.55\ \mu m$. The colormap depicts the amplitude, $A$, of converted light by the saturation and the phase, $\phi$, by the hue. (f) "Look-up table" inverting an interpolated version of (e) to specify the values of $W_y$ (saturation) and $\alpha$ (hue) required to achieve a desired $A$ and $\phi$.

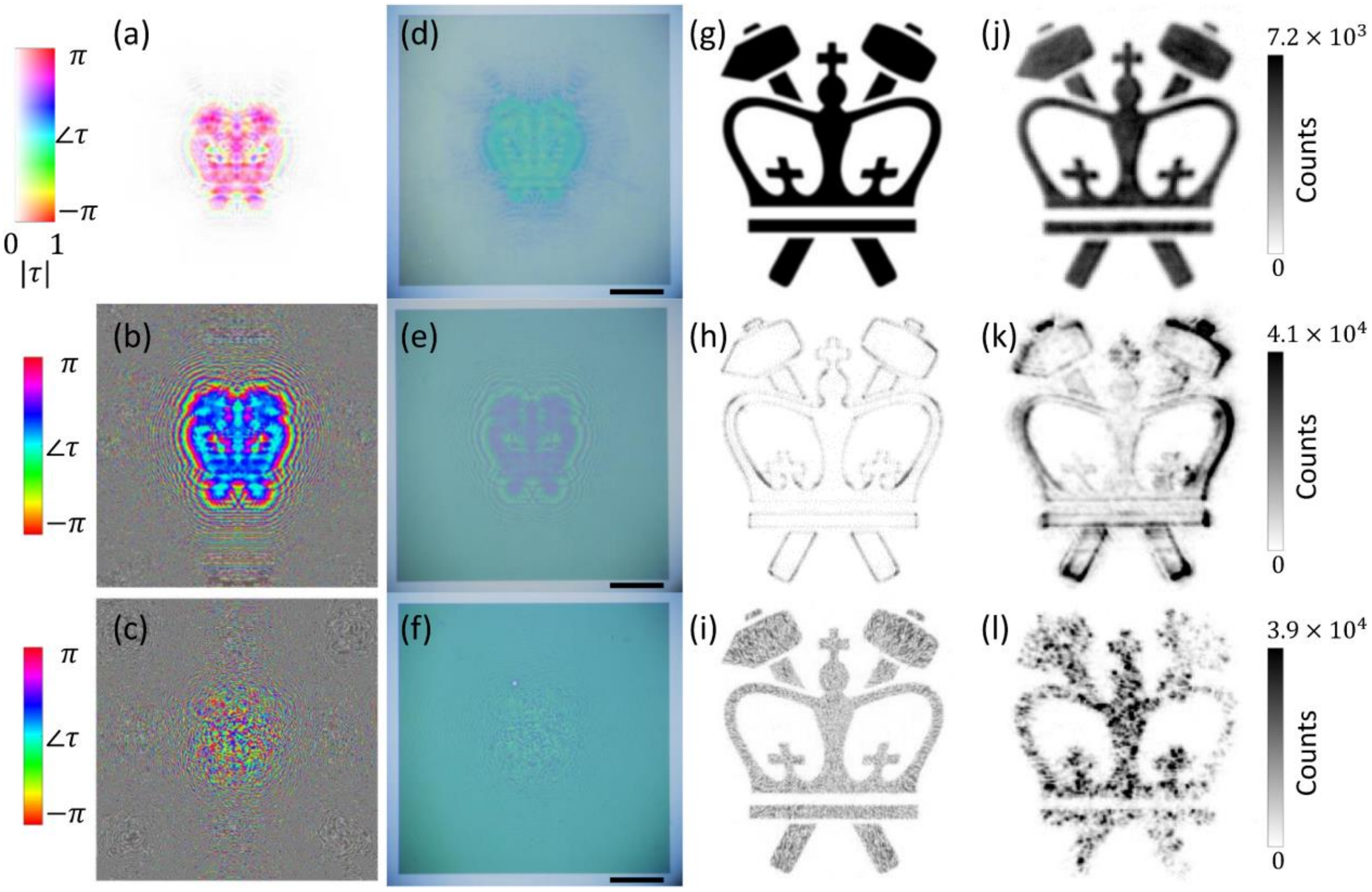

**Figure 2**. Experimental comparison of phase-amplitude (PA, top row), phase-only (PO, middle row), and Gerchberg-Saxton (GS, bottom row) holography. (a-c) The required amplitude and phase across each metasurface, where the saturation of the image corresponds to the amplitude and the hue corresponds to the phase. (d-f) Optical images of fabricated holograms. Scale bars are 150 μm. (g-i) Simulated reconstruction of the holograms. (j-l) Experimental reconstruction of the holograms, with counts shown for comparison.

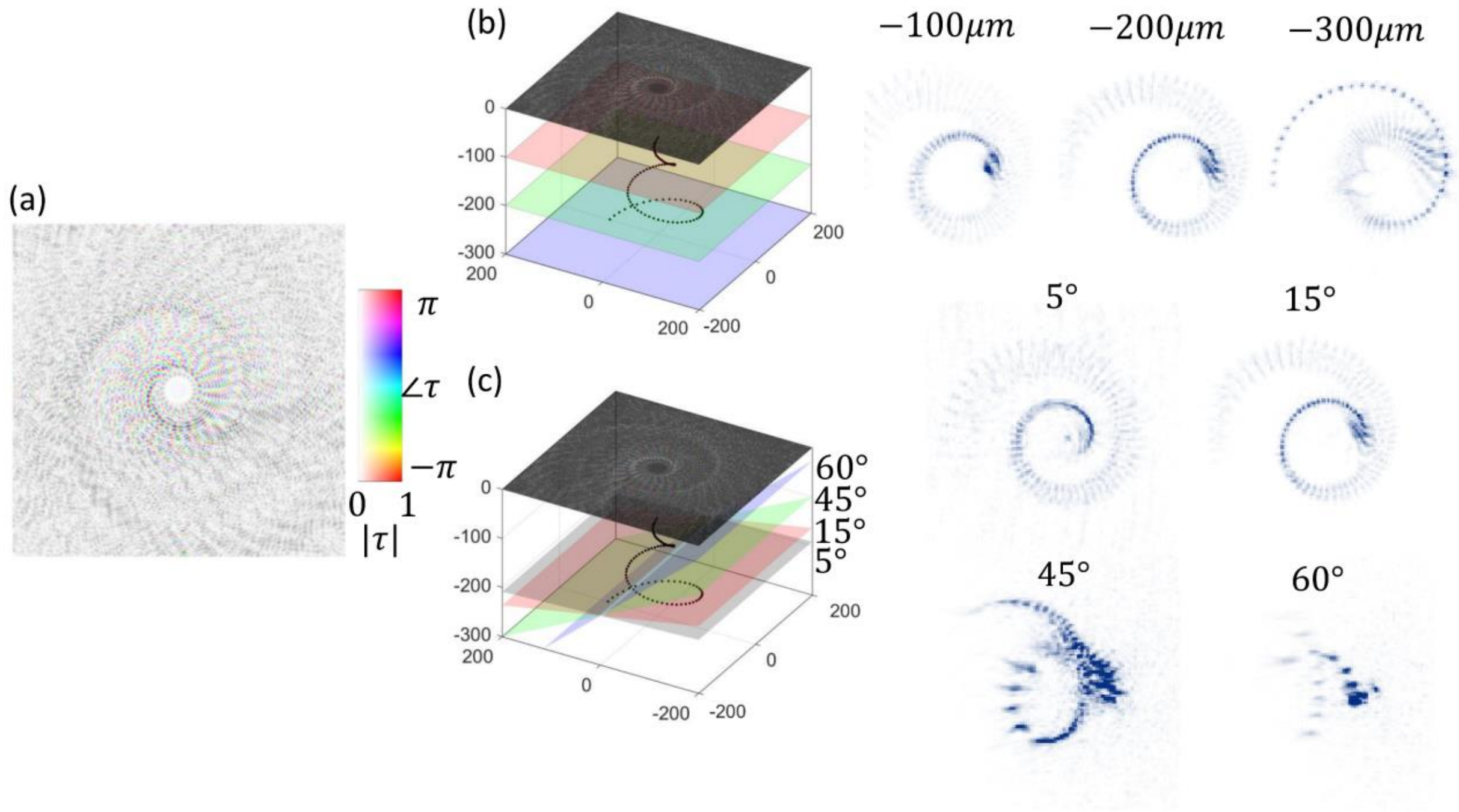

**Figure 3**. Experimental demonstration of depth and parallax in a 3D holographic object. (a) Complex transmission function, $\tau$, of a 3D coil that is 400 × 400 μm in size. (b) Experimental reconstruction of the coil at three depths, showing the 3D nature of the coil. The approximate focal plane positions relative to the metasurface plane and point sources representing the coil are shown for reference. Note that the focal planes are tilted by approximately 15° to the metasurface to reduce spurious back reflections that were present. (c) Reconstruction of the coil at varying observation angles with approximate focal planes for reference, demonstrating parallax.

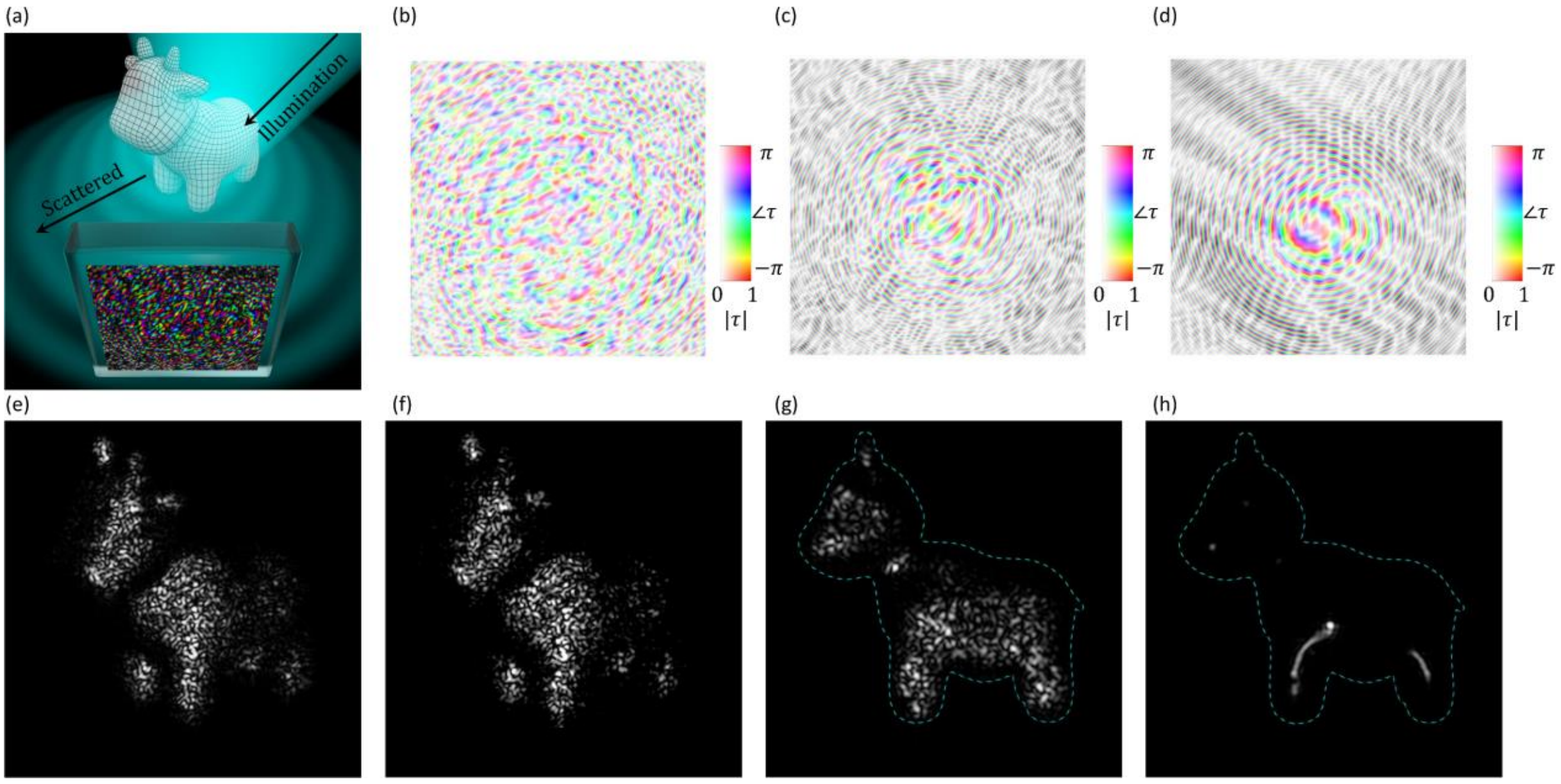

**Figure 4**. 3D computer generated holographic objects with controlled surface textures. (a) Schematic depicting the calculation of the complex transmission function, $\tau$, of a metasurface hologram to generate a complex 3D holographic object (a cow). An illuminating beam is scattered by the mesh of the cow and undergoes interference at the plane of the metasurface. (b) $\tau$ for the cow with a rough surface texture at the viewing angle shown in (e) and (f). (c) $\tau$ for the cow with a rough texture at the viewing angle shown in (g). (d) $\tau$ for the cow with a smooth texture at the viewing angle shown in (h). (e) Simulated reconstruction of the cow, showing excellent agreement with (f) the experimental reconstruction with a diode laser. (g,h) Simulated reconstructions from a different perspective, showing the effect of surface textures on the reconstruction; for the smooth cow in (h), only the specular highlights are apparent.

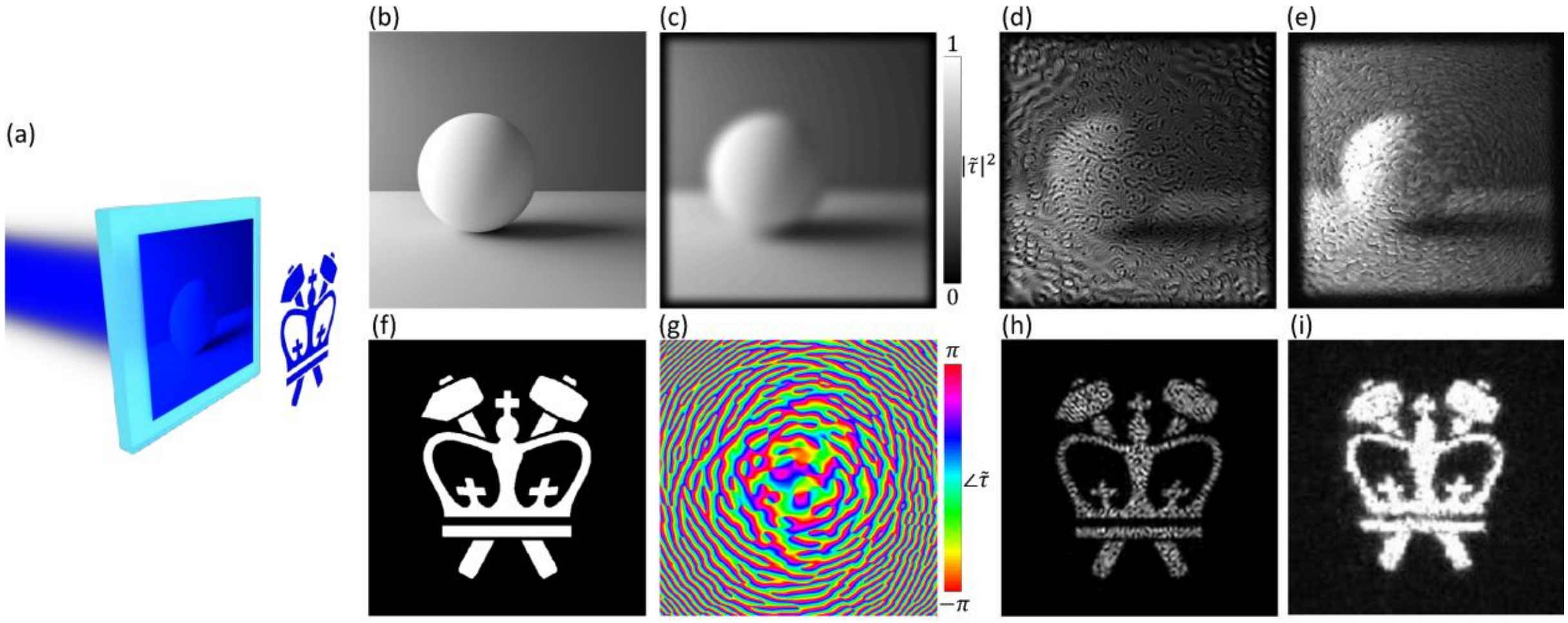

**Figure 5.** Controlling the amplitude and phase of holographic images simultaneously. (a,b) Complex transmission functions, $\tau$, of two holograms. (c,d) Simulated reconstructed complex amplitudes, $\tilde{E}$, of (a,b), yielding holographic images with identical intensity distributions but distinct phase distributions: one has a phase gradient and the other has a uniform phase. (e,f) Experimental reconstructions of two holograms corresponding to (a,b) at an observation angle of $\theta = -20°$ from the surface normal. (g,h) Experimental reconstructions of two holograms corresponding to (a,b) at an observation angle of $\theta = 0°$ from the surface normal. The dependence on observation angles is proof that the holographic images have distinct phase gradients, which correspond to distinct far-field projection angles.

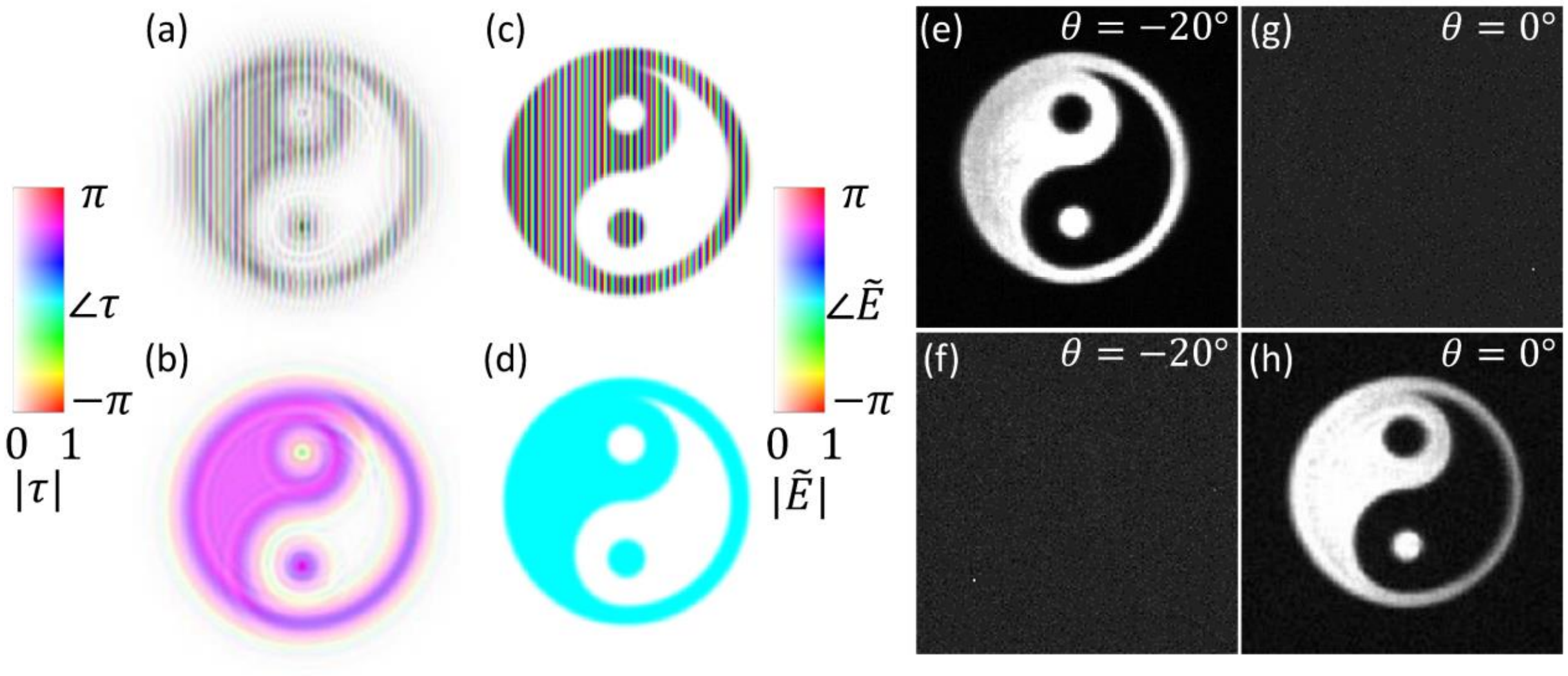

**Figure 6.** Two images encoded by a modified Gerchberg-Saxton algorithm allowing a grayscale amplitude at the metasurface plane. (a) Schematic showing the illumination of a metasurface, with an amplitude profile depicting an image of a sphere on a flat surface. The phase profile of the metasurface (not shown) encodes a holographic object (Columbia Engineering logo) at the object plane (3 mm away). (b,f) Target intensity profiles (before blurring) at the metasurface and object planes, respectively. (c,g) Intensity and phase profiles encoded on the metasurface. (d,h) Simulated

reconstructions when focused onto the metasurface and object planes, respectively. (e,i) Experimental reconstructions when focused onto the metasurface and object planes, respectively. The metasurface has side lengths of 780 $\mu m$, and the logo is approximately 250 $\mu m$ across.

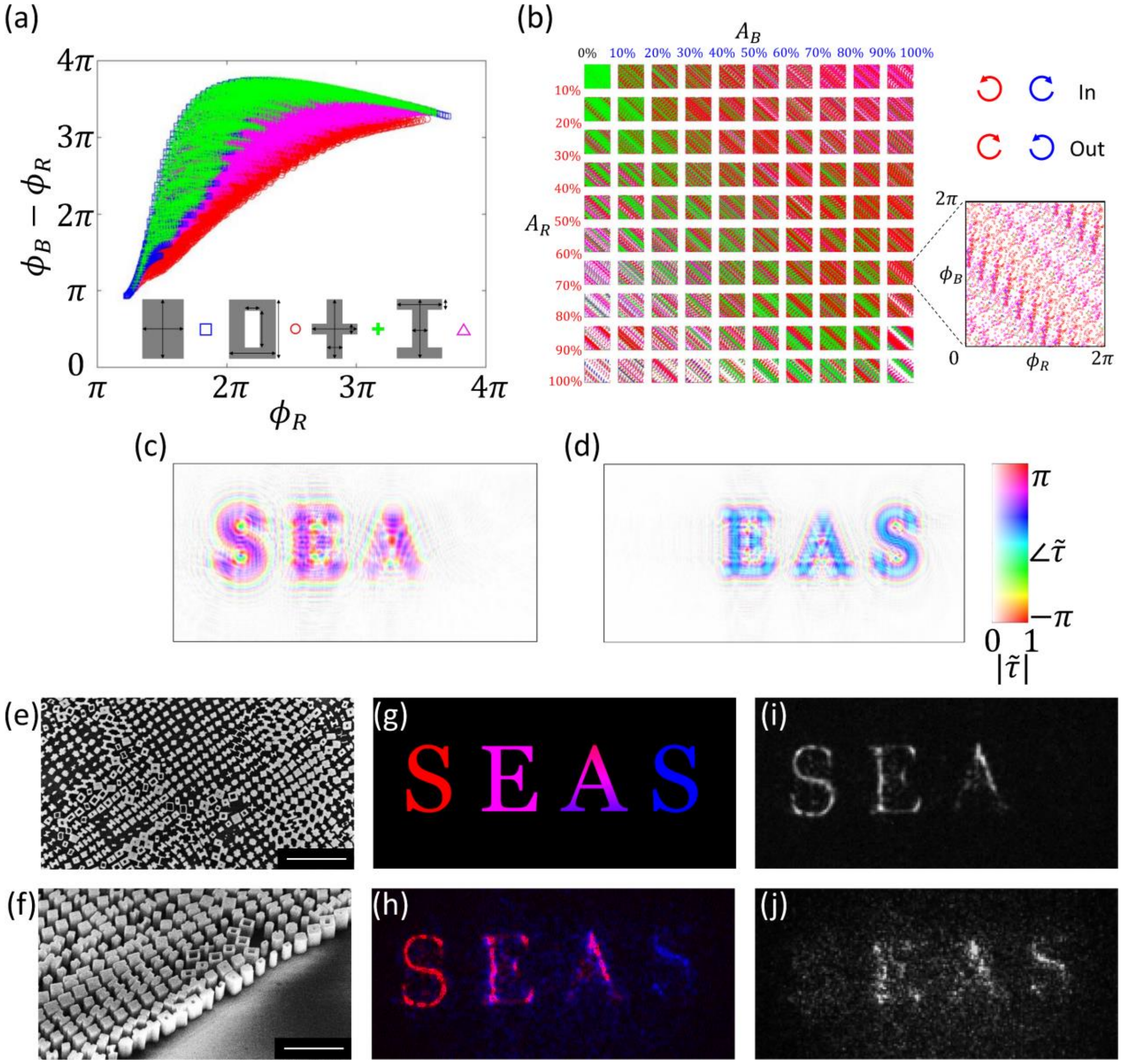

**Figure 7.** Control of the amplitude and phase at two colors simultaneously. (a) Archetypes of meta-unit cross-sections with many geometric degrees of freedom (each represented by a double-sided arrow) degenerately cover the "phase-dispersion" space of the propagation phase. (b) Visualization of the coverage of $(A_R, A_B, \phi_R, \phi_B)$ due to the meta-atoms in (a) with bins of 10% amplitude and circular polarization that is opposite for each color. (c) Complex transmission function of a two-color hologram for the red wavelength ($\lambda_{Red} = 1.65\ \mu m$). (d) Complex transmission function of a two-color hologram for the blue wavelength ($\lambda_{Blue} = 0.94\ \mu m$). (e) Scanning electron micrograph (SEM) of an example hologram, showing many instances of the archetypes from (a) with variable in-plane orientation angles. Scale bar is 3 $\mu m$. (f) SEM with a perspective view of the 1 $\mu m$-tall pillars in (e).

Scale bar is 2 $\mu m$. (g) Target two-color image. (h) Experimental reconstruction overlaying the separately measured red wavelength shown in (i) and blue wavelength shown in (j).

**Supporting Information**

**Title** Dielectric Metasurfaces for Complete and Independent Control of the Optical Amplitude and Phase

*Adam C. Overvig, Sajan Shrestha, Stephanie C. Malek, Ming Lu, Aaron Stein, Changxi Zheng, and Nanfang Yu**

**Section S1 Derivation of amplitude and phase of RCP output from a meta-atom**

Figure S1 depicts the evolution of the Jones vector through a meta-atom for the simplified case of $\alpha = 0$. To include the effects of α, we begin with incident light, $E_{inc} = |L\rangle$, coming from the substrate side, with definitions of left-hand circularly polarized light ($|L\rangle$) and right-hand circularly polarized light ($|R\rangle$) in terms of linear polarization basis, ($|X\rangle, |Y\rangle$):

$$|X\rangle = \begin{bmatrix}1\\0\end{bmatrix},$$

$$|Y\rangle = \begin{bmatrix}0\\1\end{bmatrix},$$

$$|L\rangle = \frac{1}{\sqrt{2}}(|X\rangle + \mathrm{i}\,|Y\rangle),$$

$$|R\rangle = \frac{1}{\sqrt{2}}(|X\rangle - \mathrm{i}\,|Y\rangle).$$

The state of light as a function of propagation distance $z$ through the meta-atom, $|\Psi(z)\rangle$ can be written as:

$$|\Psi(z)\rangle = \Gamma(-\alpha)M(z)\Gamma(\alpha)|L\rangle,$$

with

$$M(z) = \begin{bmatrix}A_o e^{i\phi_o(z)} & 0\\ 0 & A_e e^{i\phi_e(z)}\end{bmatrix},$$

$$\phi_o(z) = \frac{2\pi}{\lambda} n_o z,$$

$$\phi_e(z) = \frac{2\pi}{\lambda} n_e z,$$

and

$$\Gamma(\alpha) = \begin{bmatrix} cos\,(\alpha) & -sin\,(\alpha) \\ sin\,(\alpha) & cos\,(\alpha) \end{bmatrix}.$$

Taking $A_o = A_e = 1$, this becomes:

$$|\Psi(z)\rangle = \begin{bmatrix} \cos\,(\alpha) & \sin\,(\alpha) \\ -\sin\,(\alpha) & \cos\,(\alpha) \end{bmatrix} \times \begin{bmatrix} e^{i\phi_o(z)} & 0 \\ 0 & e^{i\phi_e(z)} \end{bmatrix} \times \begin{bmatrix} \cos\,(\alpha) & -\sin\,(\alpha) \\ \sin\,(\alpha) & \cos\,(\alpha) \end{bmatrix} \times \frac{1}{\sqrt{2}} \begin{bmatrix} 1 \\ i \end{bmatrix},$$

which can be simplified to:

$$|\Psi(z)\rangle = \frac{e^{i\frac{\phi_o(z)+\phi_e(z)}{2}}}{\sqrt{2}} \begin{bmatrix} \cos\left(\frac{\phi_o(z)-\phi_e(z)}{2}\right) + i\sin\left(\frac{\phi_o(z)-\phi_e(z)}{2}\right) e^{2i\alpha} \\ i\,\left(\cos\left(\frac{\phi_o(z)-\phi_e(z)}{2}\right) - i\sin\left(\frac{\phi_o(z)-\phi_e(z)}{2}\right) e^{2i\alpha}\right) \end{bmatrix}. \qquad \text{(S1)}$$

The action of the polarization filter is to select the RCP component of $|\Psi(z)\rangle$ after a propagation distance of $z = d$ (i.e., height of the meta-atom). The output from the polarization filter, $S$, is therefore calulcated by the inner product of $|R\rangle$ and $|\Psi(d)\rangle$:

$$S = \langle R|\Psi(d)\rangle = \frac{1}{\sqrt{2}}[1 \quad -i]^* \times |\Psi(z)\rangle,$$

which simplifies to equation (4) in the main text:

$$S = i\sin\left(\frac{k_0 d(n_o - n_e)}{2}\right)\exp\left(i\left(\frac{k_0 d(n_o + n_e)}{2} + 2\alpha\right)\right).$$

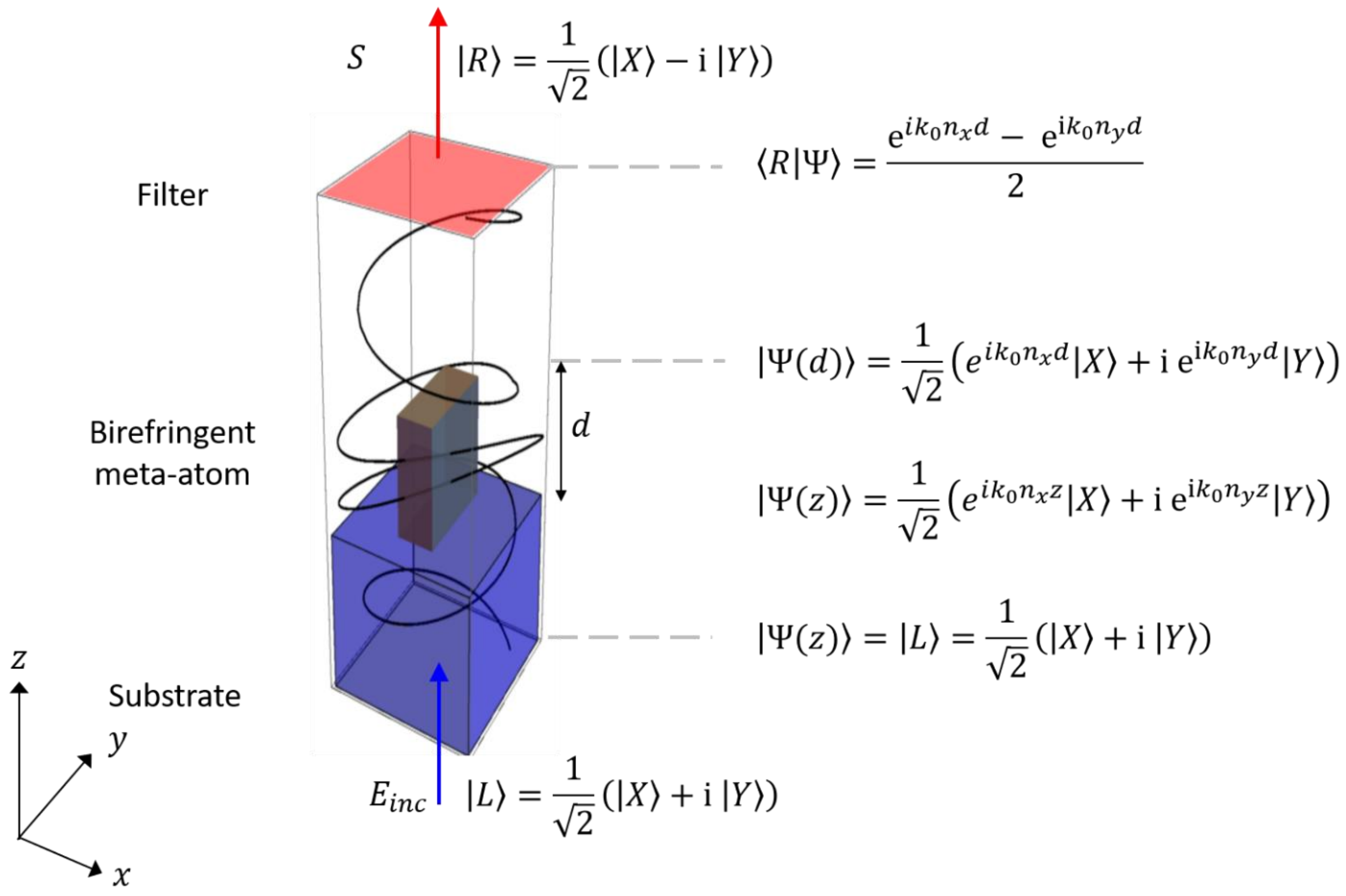

**Figure S1**. Schematic of the evolution of light through a birefringent meta-atom, with α = 0 for simplicity. LCP light is incident from the substrate side, couples into the birefringent meta-atom, evolves from LCP to a mixture of RCP and LCP (here, a complete conversion is depicted with the black curve tracing the end of the electric-field vector), and then the polarization filter selects the RCP component.

**Section S2 Meta-atom library as a polarization state converter**

We define $\eta_{conversion} = |S| = \sin\left(\frac{k_0 d(n_o - n_e)}{2}\right)$ as a measure of the birefringence of a given meta-atom. Figure S2 depicts the relationship between the output position on the Poincaré sphere and the values of $\eta_{conversion}$ and $\alpha$. The longitude, $2\psi$, and latitude, $2\chi$ of the Poincaré sphere define the two degrees of freedom determining the polarization state, and along with the intensity, $I$, are the spherical coordinates corresponding to the Stokes parameters of polarized light:

$$S_0 = I$$

$$S_1 = I\cos(2\psi)\cos(2\chi)$$

$$S_2 = I\sin(2\psi)\cos(2\chi)$$

$$S_3 = I \sin(2\chi).$$

Complete control over the output polarization state therefore requires independent control of $\psi$ and $\chi$. As depicted in Figure S2b-d, equation S1 predicts that a meta-atom library with $\eta_{conversion}$ spanning from 0 to 1, along with $\alpha$ ranging from 0 to 180°, will be able to take incident circularly polarized light (here, LCP) into any output polarization state with unity power efficiency. Full-wave simulations (seen in Figure S2e-f and detailed in Section S3) confirm this, with Figure S2e demonstrating that the efficiency can be maintained above 96% for all meta-atoms. In both cases, it is evident that independent control of $\psi$ and $\chi$ are achieved through $\alpha$ and $\eta_{conversion}$, respectively.

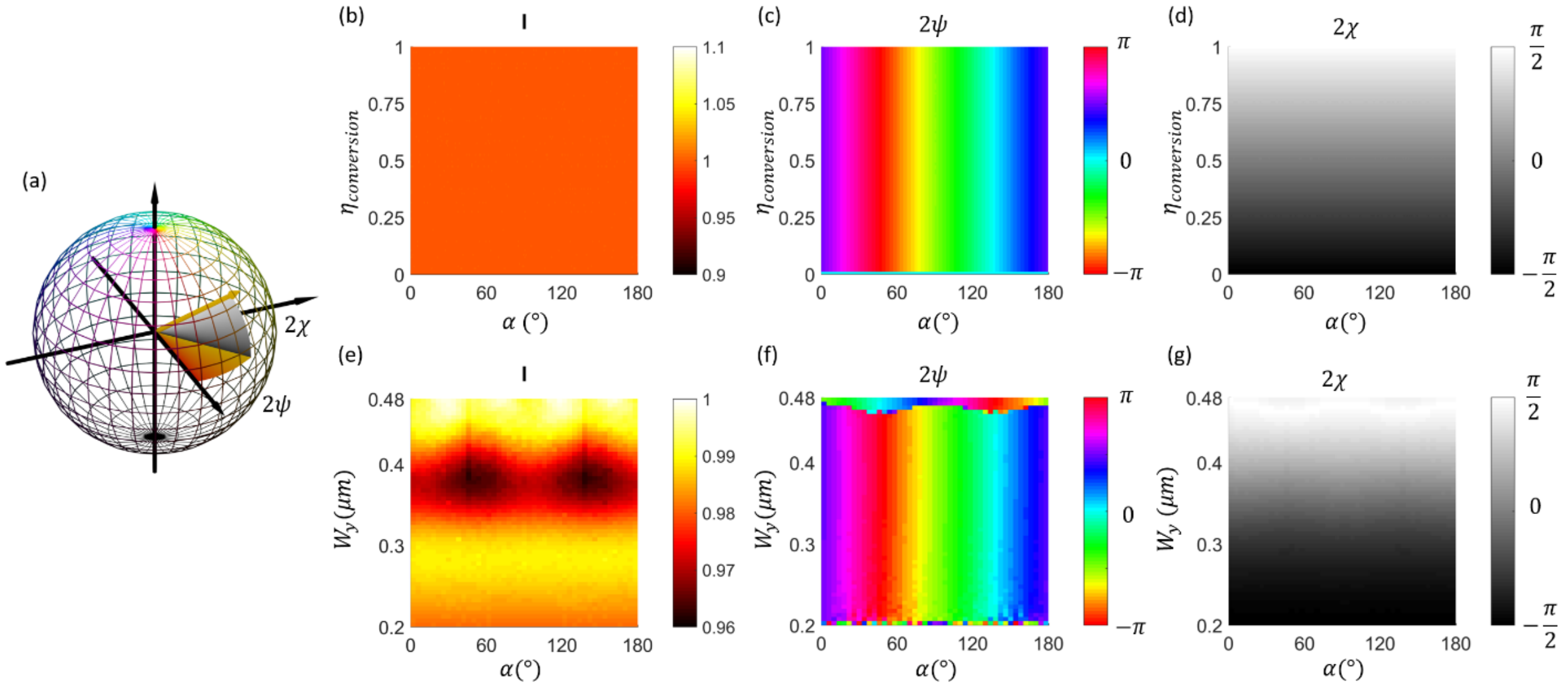

**Figure S2**. Achieving any output polarization state, visualized by the Poincaré sphere. (a) Poincaré sphere, with definitions of longitude, $2\psi$, and latitude, $2\chi$. Map of the Intensity, I (b), longitude (c), and latitude (d) predicted by equation S1, as a function of rotation angle, $\alpha$ and conversion amplitude, $\eta_{conversion} = \sin\left(\frac{k_0 d(n_o - n_e)}{2}\right)$. Map of the simulated Intensity (e), longitude (f), and latitude (g) achievable by the meta-atom library. The meta-atoms are made of amorphous silicon on fused silica substrates; the lattice constant is $P = 650\ nm$, the meta-atom height is $d = 800\ nm$, and one edge of the rectangular cross-section of the meta-atom is $W_x = 200nm$, while the other edge $W_y$ varies from $200\ nm$ to $480\ nm$; see Figure S3a for the definition of these geometric parameters. Note the complete and independent control that $\alpha$ and $W_y$ provide on longitude and latitude, respectively. Also note that the discontinuity in (f) is due to overshooting the north pole, with a difference of $\pi$ to the value of longitude.

**Section S3 Full-wave simulations of meta-atom library**

While the physical picture described in Section S1 predicts full amplitude and phase control, the precise geometric parameters capable and practical to achieve such control must be found by numerical methods. Toward this end, full-wave simulations (FDTD, Lumerical Solutions) are carried out on the individual meta-atoms, which do not require the approximations made in the description in Section S1 (most notably, that $A_o = A_e = 1$).

Figure S3a depicts the in-plane geometrical parameters to be explored numerically. The height of the meta-atoms, $d$, and the period of the lattice, $P$, are chosen to be subwavelength, but allowed to vary within that constraint. Then, with the in-plane orientation angle, $\alpha$, kept constant, the widths in the $x$ and $y$ directions, $W_x$ and $W_y$, respectively, are varied in a parameter sweep, recording the scattering (Figure S3b) and conversion efficiencies (Figure S3c). After some initial exploration, the values $d = 800\ nm$ and $P = 650\ nm$ are chosen because they not only satisfy the subwavelength condition but also yield large scattering efficiencies for a wide range of $W_x$ and $W_y$, as seen in Figure S3b. Then, a contour through this parameter space is chosen such that the conversion amplitude varies continuouslty from 0 to 1 while the scattering efficiency remains near unity (Figure S3d). Many contours could have been chosen, but for simplicity a contour with a constant value of $W_x = 200\ nm$ was chosen, allowing the contour to be characterized by $W_y$ alone.

Finally, to quantify the degree to which varying $\alpha$ changes the conversion amplitude, full-wave simulations are performed varying $\alpha$ for each value of $W_y$. The amplitude and phase of the converted light is then recorded in Figure S3e and Figure S3f, respectively. The inversion of these simulations (detailed in Seciton S4) produces a look-up table giving the required $W_y$ (Figure S3g) and $\alpha$ (Figure S3h) for a desired combination of amplitude and phase.

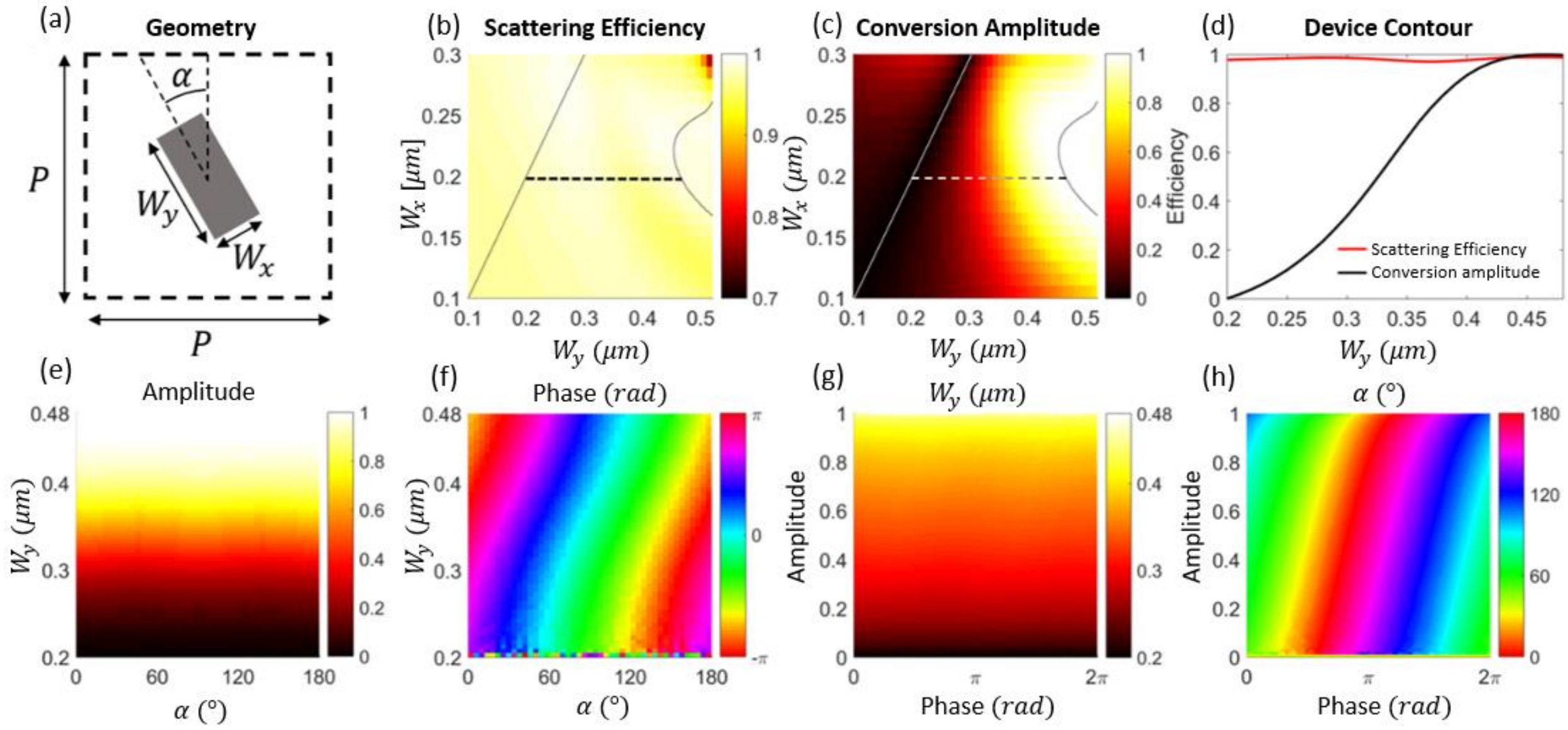

**Figure S3**. Full-wave simulations showing optical performance of the library of meta-atoms. (a) Top-view of a meta-atom showing its geometrical parameters. With $\lambda = 1.55\ \mu m$, $P = 650\ nm$, $\alpha = 0$ and the meta-atom height, $d = 800\ nm$, a range of possible values of $W_x$ and $W_y$ are swept and the forward scattering efficiency (or transmittance) (b) and conversion amplitude (from LCP to RCP) (c) are recorded. Periodic boundary conditions are assumed in the simulations. A contour representing varying $W_y$ and fixed $W_x = 200\ nm$ (dashed lines in (b) and (c)) is selected to cover the full range of conversion from LCP to RCP while maintaining high scattering efficiency (>96%) (d). With LCP incident light, $W_y$ is swept for each choice of $\alpha$ in the range of $[0°, 180°]$, and the amplitude (e) and phase (f) of output RCP light are recorded. The results of (e,f) are inverted into "look-up" tables where for a given desired combination of amplitude and phase, the required $W_y$ (g) and $\alpha$ (h) can be found. The completeness of the look-up tables demonstrates the complete and independent control over the two wavefront parameters simultaneously.

## Section S4 Look-up table construction

The process of constructing the look-up table is as follows: First, the meta-atom library simulations (Figure S3e,f) are interpolated in order to provide a library that is more continuous. This is done in lieu of additional full-wave simulations to save time, and is justified by the monotonic behavior shown in the discrete set of simulations performed. Second, a table of each combination of target phases, $\phi$, in the range of $[0, 360°)$ and amplitudes, $A$, in the range of $[0,1]$ is generated. The entries in this table take the form of a phasor: $Ae^{i\phi}$. Third, for each entry in the table, the target phasor ($A_t e^{i\phi_t}$) is compared to the achievable phasors in the interpolated meta-atom library. The geometrical parameters for the choice with minimal error is recorded along with the corresponding error ($error = |Ae^{i\phi} - A_t e^{i\phi_t}|$). The results are shown in Figure S4. Figure S4a,b depict the look-up table constructed and Figure S4c depicts the corresponding error for each entry. The maximum error is roughly 0.011 (or 1.1%).

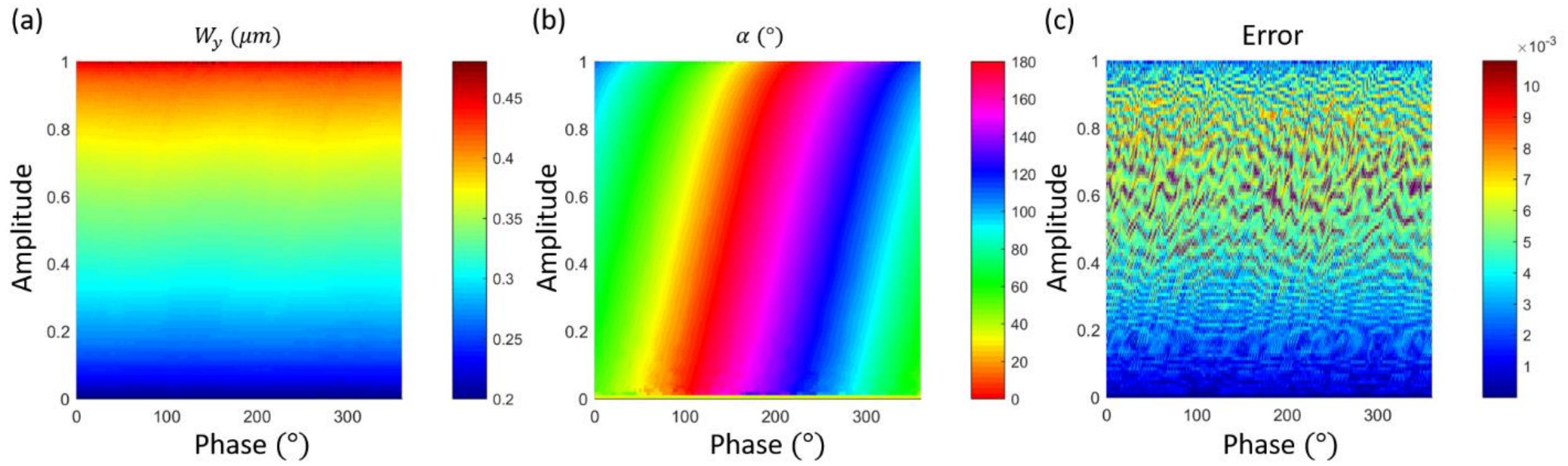

**Figure S4**. Look-up table construction. Constructed optimal choice of $W_y$ (a) and $\alpha$ (b) for each desired amplitude and phase combination. The absolute value of the difference in the target phasor and the closest achievable phasor is recorded for each target phasor in (c), showing a maximum error of 0.011, or 1.1%.

**Section S5 Effect of blur and numerical aperture on PA reconstruction**

Because free-space momentum of light is a fixed value constrained by the wavelength of light, there is an upper limit to the spatial frequencies encodable by a metasurface hologram. A useful quantification of this limit is the numerical aperture, $NA = \sin(\theta)$, where $\theta$ is a representative range of angles across which information is encoded in the hologram. As a simple case, one can consider a metasurface lens with focal spot, $f$, and diameter, $D$, as a hologram of a single point. Then, $NA = \frac{D/2}{\sqrt{f^2+(D/2)^2}}$ as usual. We can use this definition for 2D holograms where $f$ is the distance from the object plane to the metasurface plane and the diameter is replaced by the width, $W$, of the metasurface. For 3D holograms, we can take $f$ to be the shortest distance from the holographic object to the plane of the metasurface (for instance, the tip of the coil seen in Figure 3c of the main text). Generally speaking, the higher the $NA$, the smaller the features that can be resolved upon reconstruction for a given operating wavelength $\lambda$. The relevant parameters for the fabricated holograms are presented in Table S1.

**Table S1 Parameters of Fabricated Holograms**

| Hologram | $W$ ($\mu m$) | $f$ ($\mu m$) | $NA$ |
|---|---|---|---|
| Logo (Figure 2) | 750 | 750 | 0.45 |
| Coil (Figure 3) | 400 | 100 | 0.89 |
| Cow (Figure 4) | 700 | $\sim$1000 | ~0.33 |
| Yin-Yang (Figure 5) | 450 | 500 | 0.41 |
| Sphere/Logo (Figure 6) | 780 | 3000 | 0.13 |
| SEAS (Figure 7) | $400 \times 200$ | 100 | 0.89 |

Simple concepts from Fourier analysis predict that perfectly sharp boundaries in a 2D holographic image cannot be produced by a hologram with finite $NA$ because this boundary is encoded by arbitrarily large spatial frequencies. Attempting to reconstruct a perfectly sharp boundary with a finite range of spatial frequencies results in the well known phenomenon called Gibb's overshoot, in which amplitude ripples are apparent near the sharp boundary. To avoid such ripples, because of aesthetic considerations for instance, perfectly sharp boundaries should therefore be smoothed out to a degree such that the $NA$ and $\lambda$ of the experiment can faithfully encode the entire range of sptial frequencies represnted by the holographic object.

For our implementation, we apply a Gaussian blur to a target image (such as the Columbia Engineering Logo in Figure 2 of the main text, or the Yin-Yang symbol in Figure S5) to elimiate the presence of Gibb's overshoot. A numerical exploration of the visual impact of a Gaussian blur with characteristic size of $b$ pixels (implemented by the Matlab function imgaussfilt($Image$, $b$) and with the physical size of a pixel being the same as the lattice spacing $P$ of the hologram) is seen in Figure S5. The metasurface is $W = 400\ \mu m$ in width and the object is placed at varying planes a distance $f$ away. The operating wavelength is $\lambda = 1.55\ \mu m$ As described above, the $NA$ is then calculated according to $NA = \frac{W/2}{\sqrt{f^2+(W/2)^2}}$.

It is apparent from Figure S5 that for for higher $NA$, less blur (smaller $b$) is needed to remove the overshoot, consistent with the fact that a higher $NA$ metasurface encodes a wider range of spatial frequencies. However, due to the sampling theorem, a metasurface with a finite lattice spacing faces an upper limit of the value of $NA$ achieveable (beyond which a metasurface behaves like a conventional grating), resulting in a degradation in image quality regardless of the degree of blur (bottom row of Figure S5). Alternatively, a larger ratio $W/\lambda$ can be used to achieve the same image improvement without increasing the $NA$. Considering practical constraints of nanofabrication, we use metasurface dimensions less than $W = 1\ mm$, and correspondingly use the process depicted in Figure S5 to guide the choice of $NA$ (reported in Table S1) to produce aesthetically pleasing results for the Columbia Engineering Logo seen in Figure 2 of the main text.

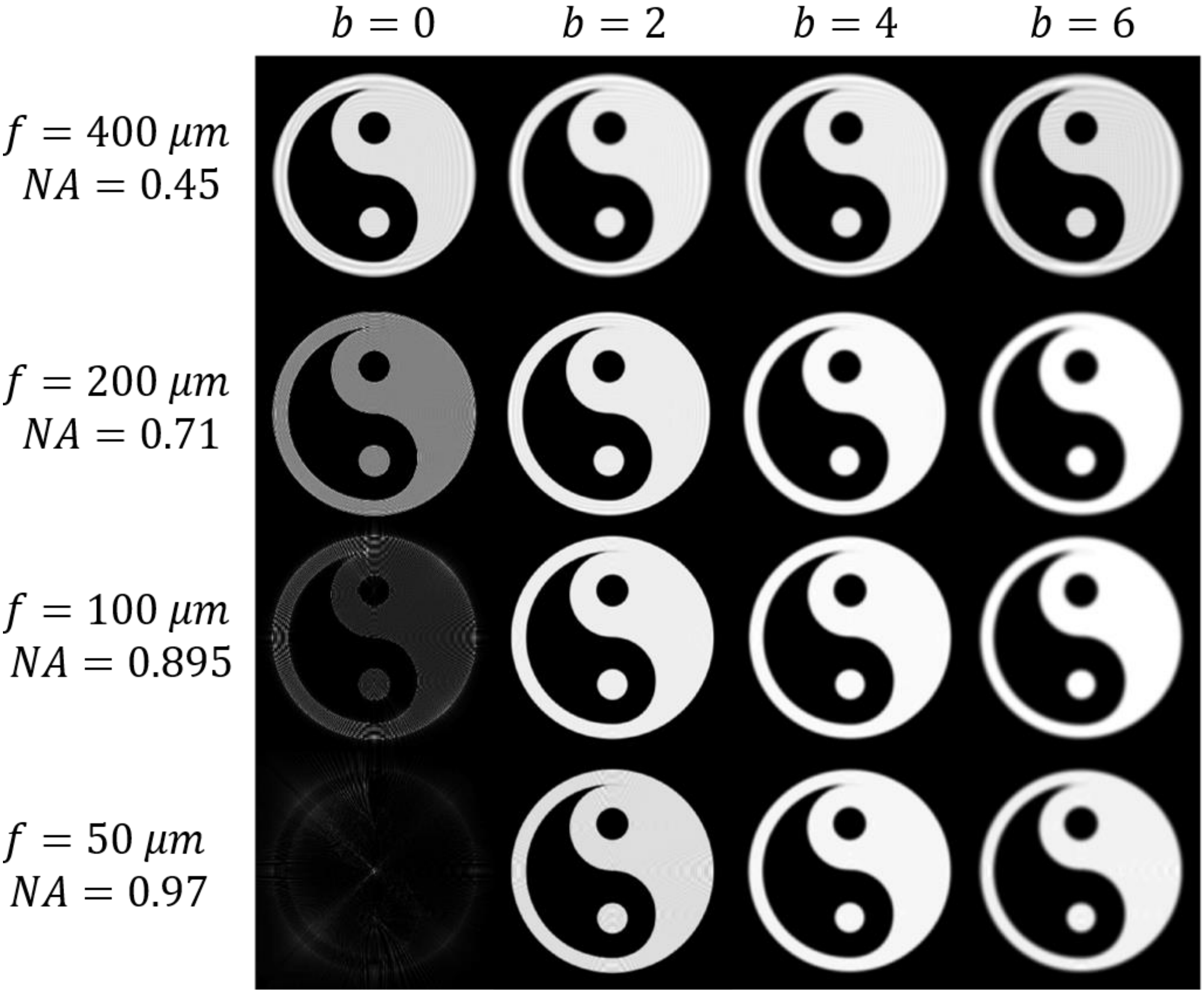

**Figure S5**. Numerical reconstruction at various combinations of numerical aperture $NA$ and degree of Gaussin blur (with characteristic size of $b$ pixels). Gibb's overshoot reduces as $b$ increases, and the magnitude of $b$ required to eliminate overshoot reduces as $NA$ increases. However, past a certain value of $NA$, the image quality degrades due to the insufficient sampling of the metasurface (due its finite lattice spacing, $P$). Careful choice is therefore required of $NA$ and $b$ such that overshoot is reduced, the image isn't too visibly blurry, and the image is not degraded due to insufficient sampling.

## Section S6 Fabrication

The fabrication process is summarized in Figure S6. A fused silica wafer is cleaned (with successive acetone, isopropyl alcohol (IPA), and deionized water (DIW) rinses, followed by dry nitrogen gun) in preparation of amorphous silicon growth. The amorphous silicon is grown to a thickness of 800 nm by chemical vapor deposition at a temperature of 200°C. The wafer is protected by a layer of poly(methyl methacrylate) (PMMA) spun on and baked at 180°C for 5 minutes. The wafer is cleaved into smaller pieces (roughly 1 cm ×2 cm in dimension). The protective layer is removed by an identical cleaning process as above, and replaced by a double layer of PMMA. The first layer has molecular weight of 496,000 and a dilution of 4% in anisole. The second (top) layer has molecular weight of 950,000 and a dilution of 2% in anisole. Both are spun at 4000 rpm and baked at 180°C. The first layer is baked for 10 minutes, and the second for 2 minutes.

Next, the hologram patterns are written by electron beam lithography (JEOL 6300) at a beam energy of 100 keV, beam current of 500 pA, and with a base dose of 740 $\mu C/cm^2$ and appropriate proximity effect corrections (BEAMER). The resulting patterns are developed in a solution of 3:1 IPA:DIW for 2 minutes in a cold bath set at 5°C and then rinsed for 30 seconds in DIW at room temperature to stop development. A dry nitrogen gun is used to lightly remove remaining water from the samples.

The exposed and developed samples are then placed in a physical evaporator (LESKER) to deposit roughly 15 nm of aluminum oxide by electron beam evaporation. Lift-off is performed by dissolution of the remaining resist in N-Methyl-2-pyrrolidone (NMP) at 85°C for 4 hours. The sample is then transferred to an acetone bath and sonicated for 5 seconds to aid the completion of lift-off. After a final rinse in IPA, dry nitrogen is blown to dry the samples.

Finally, the pattern is transferred from the aluminum oxide mask to the amorphous silicon by dry etching (Oxford). The sample is attached to a silicon carrier wafer by vacuum grease (to ensure good thermal contact during etching) and placed in the etching chamber. A combination of $SF_6$ and $O_2$ gases, and inductively coupled plasma power and RF power are used to control the etch rate and sidewall slope. The temperature is held at -100°C for improved sidewall smoothness.

The vacuum grease is removed by careful application of acetone and IPA by a cleanroom wipe. Light drying with a nitrogen gun finishes the removal of the vacuum grease from the back of the wafer.

The aluminum dioxide mask is left on because its very small thickness and dielectric nature make the optical impact of its presence negligible. Removal could be achieved by soaking in ammonium hydroxide, preferentially dissolving it without affecting the silicon or fused silica wafer.

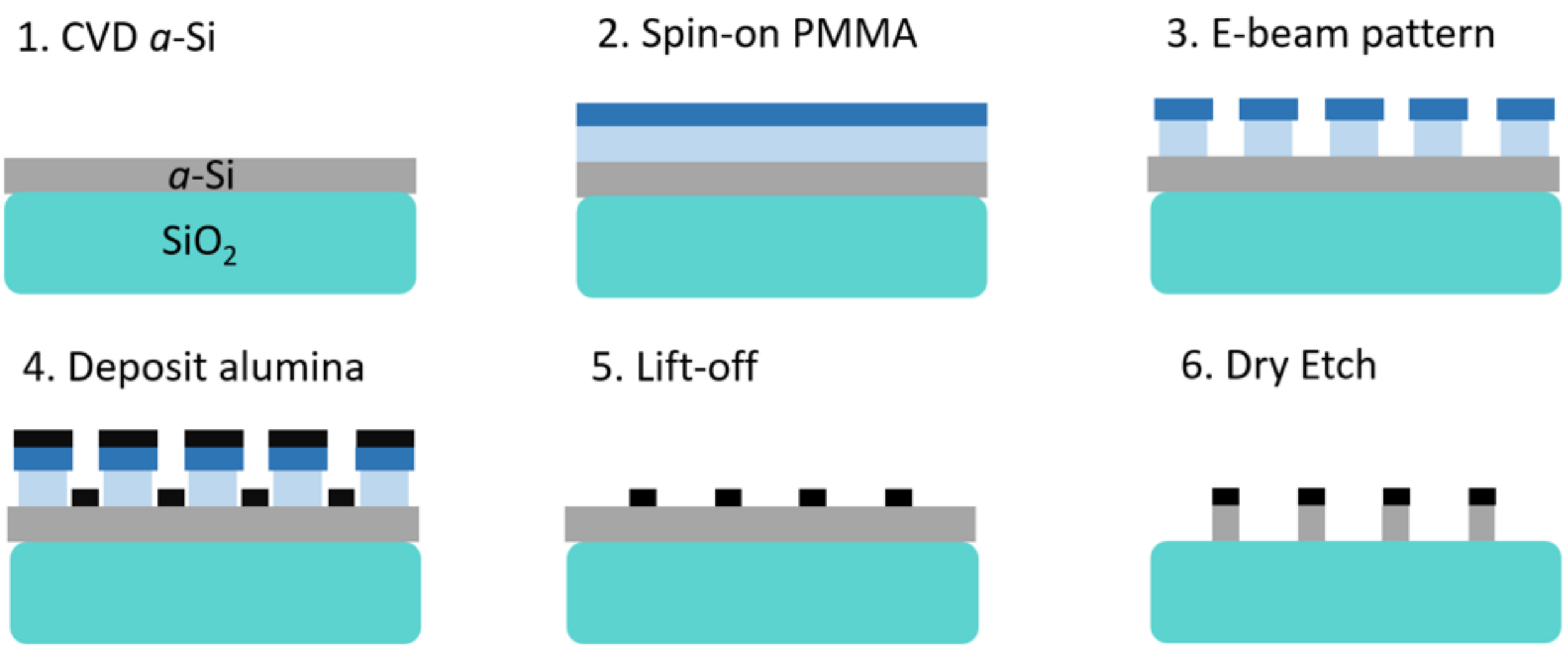

**Figure S6**. Fabrication process flow. 1. Chemical vapor deposition (CVD) of amorphous silicon (*a*-Si) on a clean fused silica wafer. 2. Spinning of double-layer PMMA electron-beam resist layer. 3. Exposure by electron-beam lithography tool and development in 3:1 IPA/DIW solution at 5°C. 4. Electron-beam deposition of alumina. 5. Chemical dissolution of remaining

resist, lifting-off unwanted alumina. 6. ICP etching transferring the alumina mask pattern into the $a$-Si layer.

**Section S7 Optical characterization set-up**

Figure S7 schematically depicts the setup used for experimental reconstruction of holographic scenes by our metasurface holograms. A set of collimating optics passes circularly polarized light to the metasurface. Light is collected and analyzed by the observation optics. The observation optics and collimating optics are linked by a swivel mount allowing a varying angle, θ, between the two. Due to the weight of the near-infrared (NIR) camera (Nirvana InGaAs camera, Princeton Instruments), the observation optics is stationary and the collimating optics are moved to change θ. The metasurface is aligned to the axis of rotation of the swivel mount by an $(x, y, z)$ dovetail stage system attached to the collimating optics. In this way, when θ is changed, the illumination condition is fixed.

The collimating optics include a fiber collimator passing input laser light from a tunable laser source to a redirecting mirror and then to a circular polarizer before finally illuminating the metasurface from the substrate side. These collimating optics are all linked together in a cage system (cage parts are omitted for clarity in Figure S7) to the swivel mount. The metasurface is mounted on a rotation mount for control of an additional Euler angle, $\phi$.

The observation setup includes an infinity-corrected 10× objective, which collects light scattered by the metasurface, and passes it through a tube lens.Then a polarization filter and iris are used to help reduce unwanted light from reaching the camera sensor.

Note that the circular polarizer and polarization filter are identical optical elements but with opposite chirality and orientation; they are composed of a polymer polarizer cemented to a polymer quarter waveplate aligned at a ±45° angle to the fast axis of the waveplate. Light incident on the circular polarizer hits the polarizer side first, and then the resulting linearly polarized light is converted by the quarter waveplate into circularly polarized light, regardless of the polarization outputted by the fiber collimator. The “polarization filter” is the the opposite handedness of the circular polarizer, and oriented such that the quarter waveplate is illuminated first. Light of the opposite handedness than that created by the circular polarizer is therefore converted by the quarter waveplate to linearly polarized light that passes through the polarizer side, while light with the same handedness is converted by the quarter waveplate to the orthogonal linear polarization, which is absorbed by the polarizer.

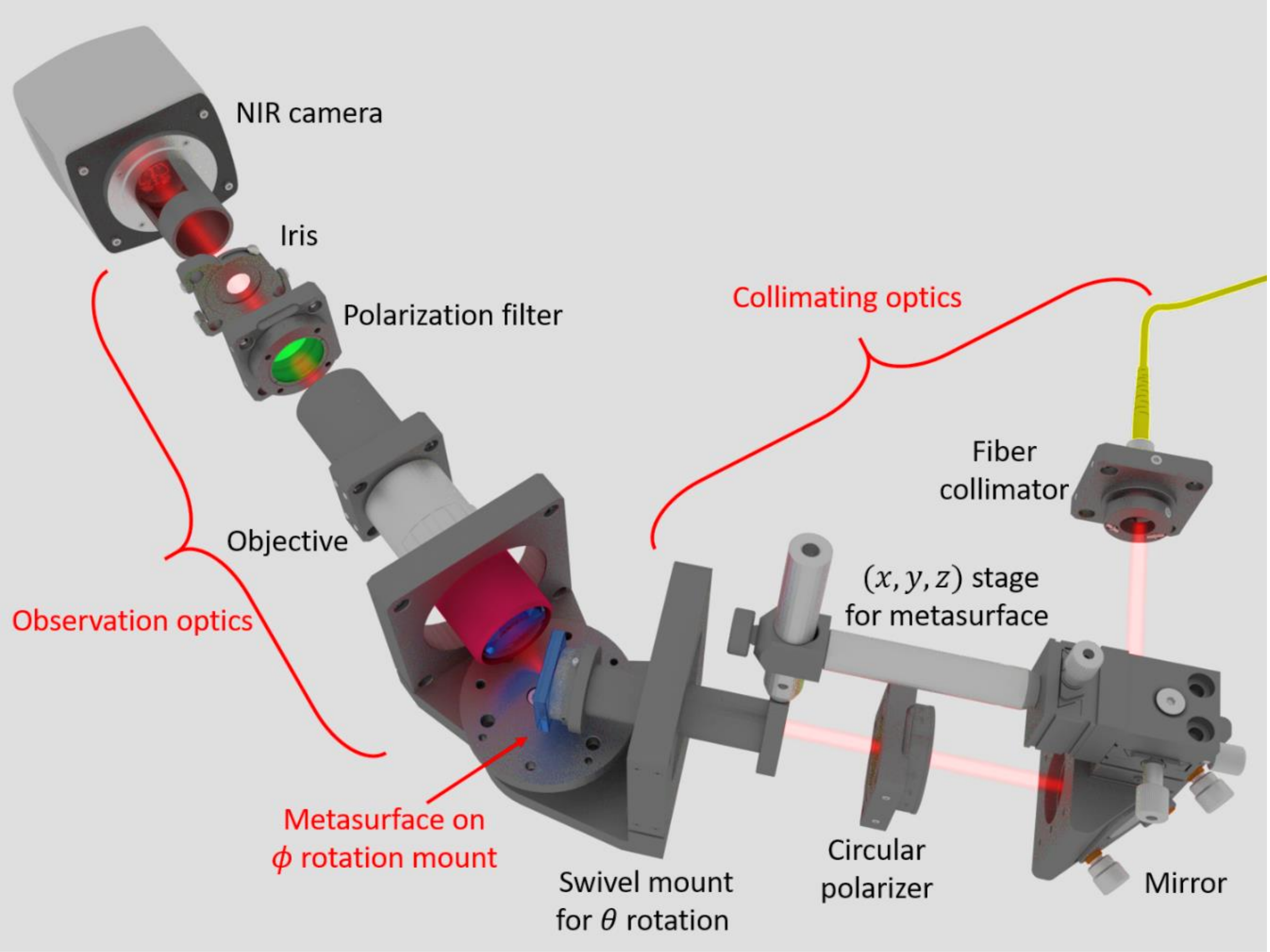

**Figure S7.** Schematic of optical setup for optical reconstruction of holographic scenes at various observation angles. Cage system parts are omitted for schematic clarity, but serve to keep the collimating condition of the light incident on the metasurface constant for varying swivel angles, $\theta$.

## Section S8 Wavelength dependence of 2D holograms

To test the dependence on wavelength of the experimental reconstruction of 2D holographic images, light generated by a supercontinuum source (NKT Photonics) is passed through a monochromator (Horiba) and then passed to the optical setup with an optical fiber. The rest of the experiment is as depicted above. Note that the circular polarizer (ThorLabs) is designed for the operating wavelength of 1,500 nm, and has roughly 4% error in phase retardation at 1,500 nm and 1,600 nm and 8% error at 1,450 nm, which may contribute to the degradation of the holographic images slightly. A wavelength of 1,650 nm is beyond the bandwidth of the fiber used for this experiment. Notwithstanding the contributions of these errors, the bandwidth of the metasurface holograms is evidently comparable to the well-known broadband behavior of metasurfaces based on the geometric phase, as shown in Figure S8. Images are as recorded, without flipping the logo horizontally as done for the main text (to match the desired orientation).

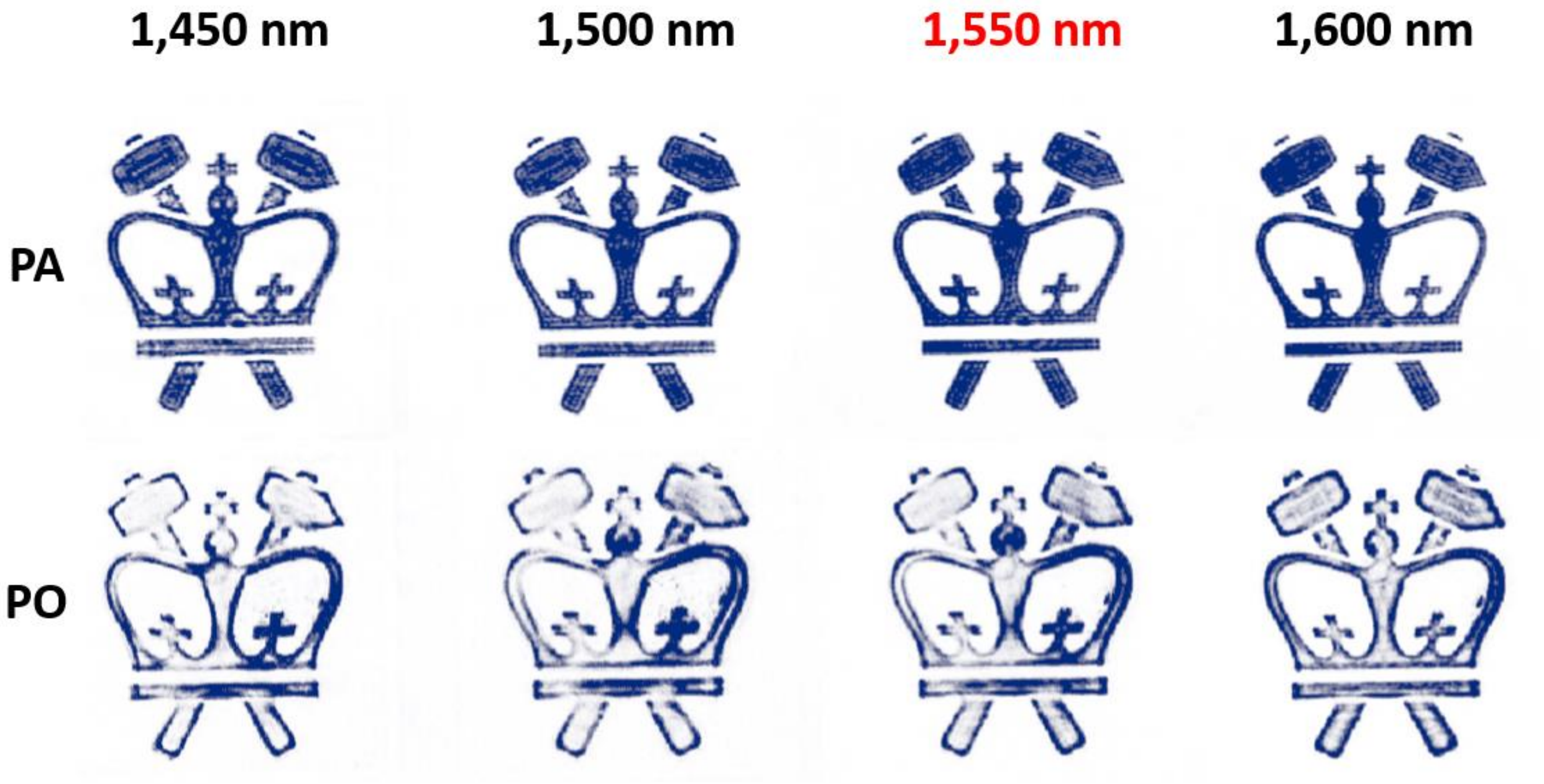

**Figure S8**. Wavelength dependence of 2D holography comparing phase and amplitude (PA, top row) to phase only (PO, bottom row) holograms for four selected wavelengths. Design wavelength of 1,550 nm is highlighted in red, and the overall bandwidth explored (150 nm) is greater than the typical of an LED centered at the operating wavelength.

**Section S9 Computer generation of the 3D hologram**

To generate the 3D hologram, we set a virtual scene wherein the cow is illuminated by an incoming plane wave. We place a hologram plane in front of the cow, and compute at every hologram pixel the optical phase and amplitude, which is a superposition of light waves reflected by the cow's surface region that is not occluded from the incident light. We compute the phase and amplitude at each hologram pixel using Monte Carlo integration over the cow mesh: we sample points over the surface mesh, and sum the complex electric field contributed by a point source located at each sampled point. In order to account for the rough surface of the cow, we also randomly perturb the phase delay between each surface point and the pixel position. The output of this simulation process is a 2D array of complex numbers, describing the phase and amplitude distribution over the hologram.

**Section S10 Simulation of optical reconstruction**

Computer reconstruction of the 3D holographic cow mimics image formation in the eye or in a camera. We treat the CGH as an input "transparency" placed directly behind a virtual lens with a focal distance of 2.45 mm. The image plane is 9.8 mm away from this virtual lens, bringing into focus the front of the cow (which is 0.5 mm in its largest dimension), which is centered roughly 5 mm behind the metasurface. In this simulation setup, the CGH serves as a spatial light modulator that shapes the phase and amplitude of the output light field at every of its pixels. We then compute the light field intensity received on an imaging plane placed in front of the lens. The imaging plane is selected to be near the head of the cow. The simulation setup enables a fast computation of the light intensity on the imaging plane using Fourier transformation.

**Section S11 Experimental reconstruction with varying coherence**

To study the impact of coherence on the optical reconstruction, we modify the optical setup to include a light emitting diode (LED) in place of the lasers. As depicted in Figure S9, an iris is added between the LED and the metasurface (labelled MS) to allow a varied degree of spatial incoherence. When the iris is almost closed, the source is approximately spatially coherent, with a temporal coherence limited by the bandwidth of the LED (roughly $\Delta\lambda$=120 nm). The reconstruction in this case is comparable to reconstruction using the diode laser (Figure 3g in the main text), but slightly blurred due to the temporal incoherence. As the iris is opened, spatial incoherence adds to this blur.

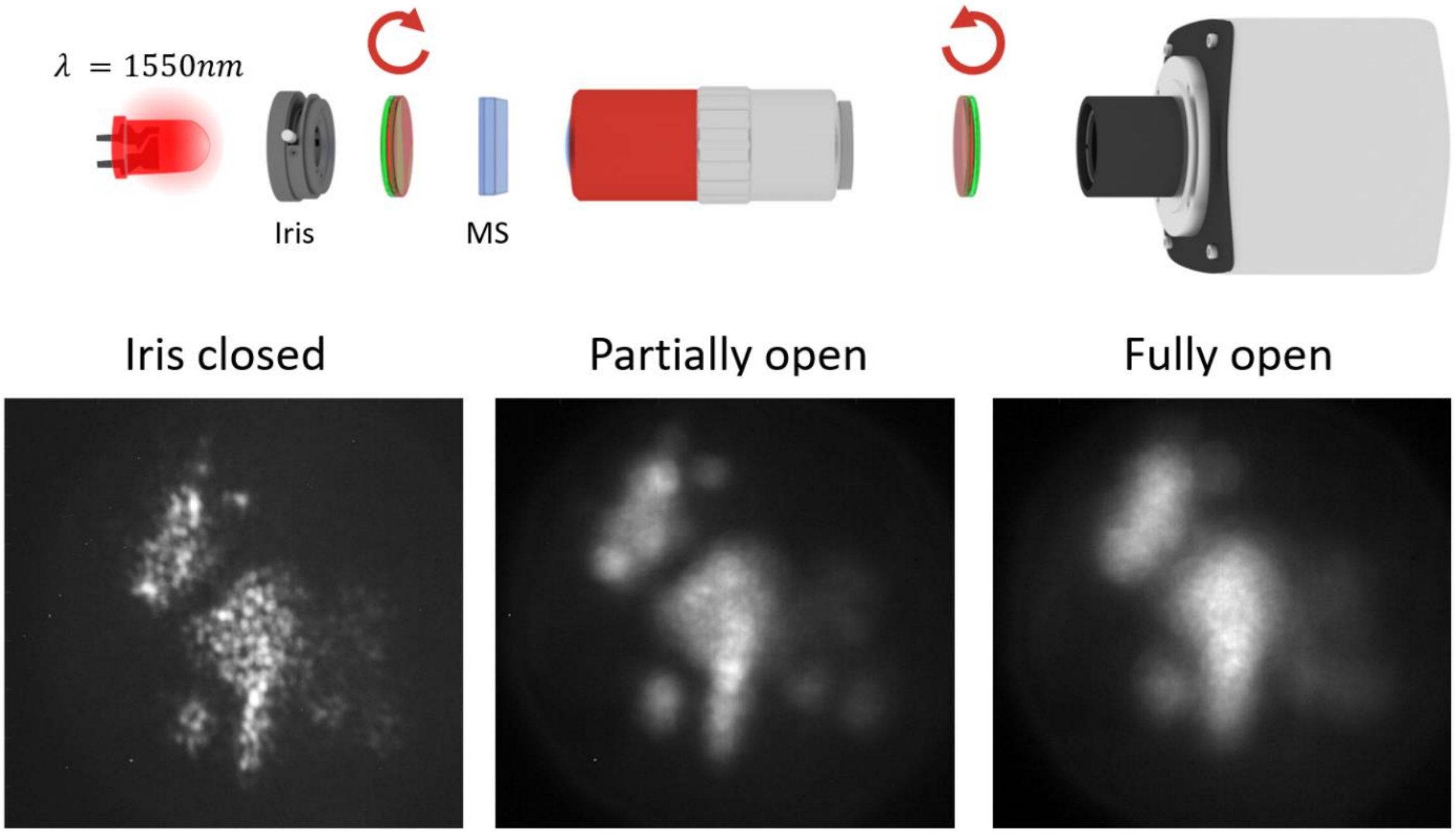

**Figure S9**. Experimental reconstruction of the cow using an LED. (Top) Schematic of the experiment using an LED and an iris for reconstruction. (Bottom) Various opening sizes of the iris yield differing degrees of spatial incoherence, resulting in a reduction in speckle as the iris is gradually opened.

**Section S12 Discussion on efficiency of holography**

It is natural to inquire as to the efficiency trade-offs between PO and PA holography, where we consider efficiency as defined by the amount of power contributing to the final image divided by the power incident on the metasurface. Of course, by design PA holography will necessarily use less of the input power than PO holography. But how much less power used is not easily generalized. It is highly case-dependent, depending on (1) the target intensity distribution of the holographic object in question, (2) the illumination pattern (e.g., shape of incident beam), and (3) the numerical aperture of the metasurface.

In particular, there is a trade-off between how much of the incident light is used and the magnitude of the ringing artifacts present upon reconstruction. Figure S10 depicts a simple case demonstrating this trade-off. A simple 1D holographic image of a blurred step function a distance $f = 250\ \mu m$ away from the metasurface is numerically reconstructed for metasurfaces of varying width, $W$. As $W$ increases, the amount of spatial frequencies encoded in the metasurface increases, and so the better the fidelity of the object upon reconstruction becomes (measured here by the root-mean-square (RMS) error compared to the target profile). However, it is apparent that this comes from extending an ever-decreasing tail of amplitude at the metasurface plane, meaning that normalized to the incident power (assumed here to be top-hat excitation with a width of $W$), the efficiency is dropping. Note how the efficiency monotonically decreases, while the RMS error generally trends downwards as well. A choice of RMS error must be made such that the reconstructed object will be considered of sufficiently high fidelity. This choice directly impacts the resulting efficiency, making the efficiency dependent on the quality of the holographic image, and therefore ambiguous in comparison to the case of a PO hologram.

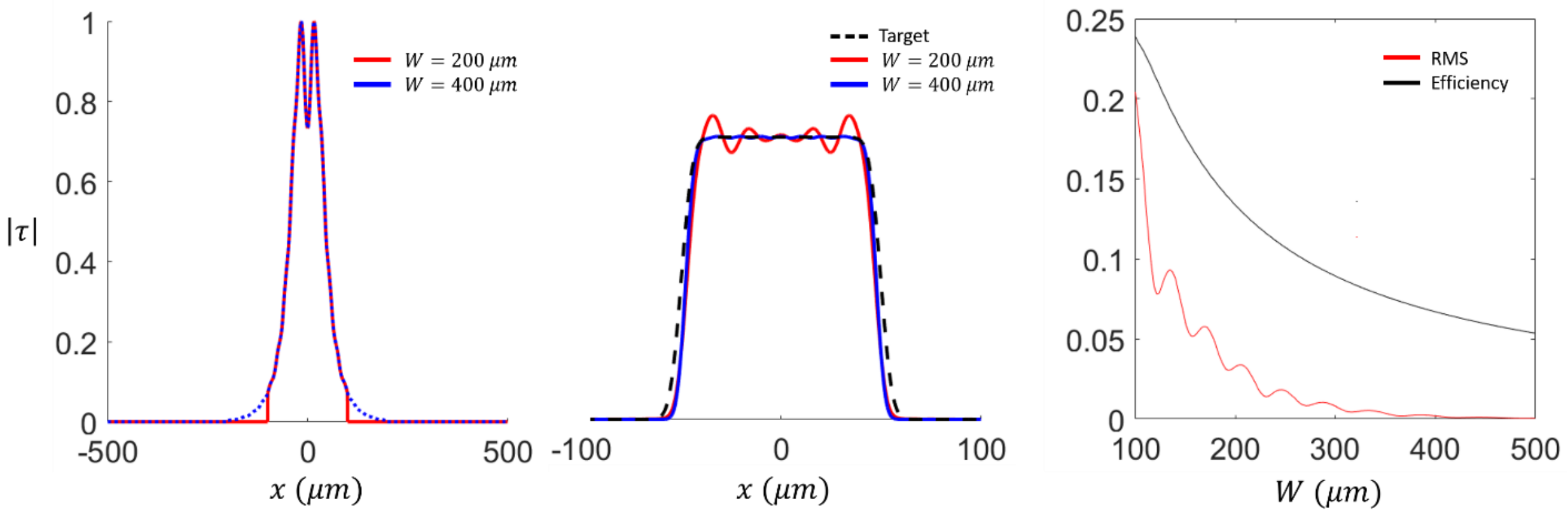

**Figure S10**. Trade-offs between Efficiency and image quality (RMS error). (left) Amplitude distribution for two example metasurfaces of different widths, $W$ (note the abrupt cutoff for the smaller metasurface). (middle) Intensity profiles of the reconstructed holographic images $250\ \mu m$ away from the metasurface plane, showing worse ringing artifacts for the smaller metasurface. (right) RMS error and Efficiency as a function of $W$, showing that the amount of incident light (assumed to be top-hat excitation with lateral extent $W$) being used decreases, but the RMS error also decreases.

**Section S13 Comparison between PA, PO, and AO holography**

When only a single degree of freedom is controllable, it is generally well-known that phase is more useful than amplitude. Here, we briefly explore and compare holography using a library of meta-atoms with phase-amplitude (PA) control (with no GS algorithm) to holography using two sub-libraries, one with phase-only (PO) control and the other with amplitude-only (AO) control. In both of the latter cases, the lack of control over both phase and amplitude simultaneously requires a GS algorithm to create holograms. Figure S11 shows the results of the three cases, demonstrating that while AO control is capable of producing an image resembling the target object, the PO case is significantly improved. This confirms the presupposition that phase is more important than amplitude, and supports the interpretation in the main text that the role of amplitude is to correctly weigh the spatial frequencies of light waves produced by the metasurface: in PO, all spatial frequencies are present, and only their relative phases can be tuned; in AO, the spatial frequencie present are tuned, but their phases are all equal; in PA, the relative phases and amplitudes of all spatial frequencies are modulated.

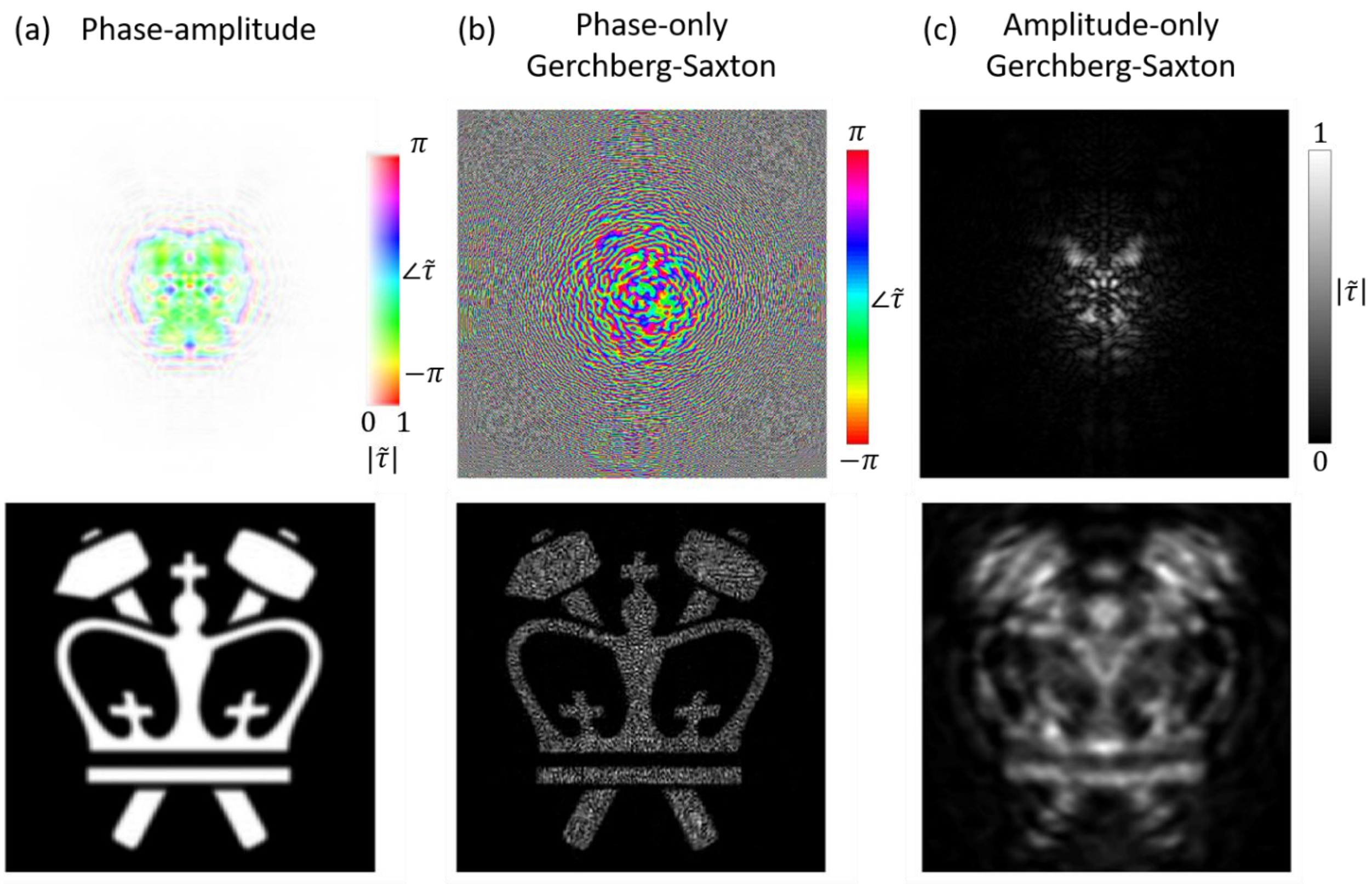

**Figure S11**. Comparison between phase-amplitude (a), phase-only Gerchberg-Saxton (b), and amplitude-only Gerchberg-Saxton (c) holography. Top row contains the metasurface complex transmission function, and the bottom row contains the simulated reconstructions. Note the degredation of the fidelity of the holographic images from left to right.

**Section S14 Gerchberg-Saxton with Phase-Amplitude control**

We briefly numerically and experimentally explore the trade-offs in the image quality at the metasurface and object planes using the GS algorithm modifed to allow a grayscale intensity mask at the metasurface plane. Figure S12 shows that as the object plane becomes closer to the metasurface plane (a distance $f$ away from the metasurface plane), the holographic image (the Columbia Engineering Logo) improves, but the image of the metasurface itself degrades. Conversely, at large $f$, the metasurface image is much improved, but the holographic image is severely degraded.

This dependence can easily be understood by considering the varying numerical aperture of the system. As $f$ decreases, the numerical aperture of the metasurface (that is, the range of spatial frequencies of the holographic object that are encoded by the metasurface) grows, meaning the object image's quality improves. However, as $f$ reduces, the required spatial frequency of the phase variance grows. Consequently, for a small region on the metasurface plane, the phase may vary rapidly while the amplitude varies slowly, even containing phase discontinuities or sigularities. Coherently imaging such a complex field will generally yield a highly speckly image due to the destructive intereference of adjacent pixels. This destructive interference due to phase variance can be seen most clearly at large $f$, where phase varies slowly, and in only a handful of locations are there sigularities. The correspondence of such sigularities and the dark artifacts can be closely correlated by comparison across the bottom row of Figure S12. The phase sigularities generally vary across $2\pi$ around a contour circling a singualrity, while the amplitude varies slowly along the same contours. This leads to destructive interference at the (simulated) camera plane.

Two example cases from Figure S12 are implemented experimentally and shown in Figure S13. As expected, the hologram with larger $f$ has better image quality at the metasurface plane while the hologram with smaller $f$ has better image quality at the object plane. Lastly, a binary image (a Yin-Yang symbol) at the metasurface plane is shown in Figure S13c but with the same holographic object at the object plane.

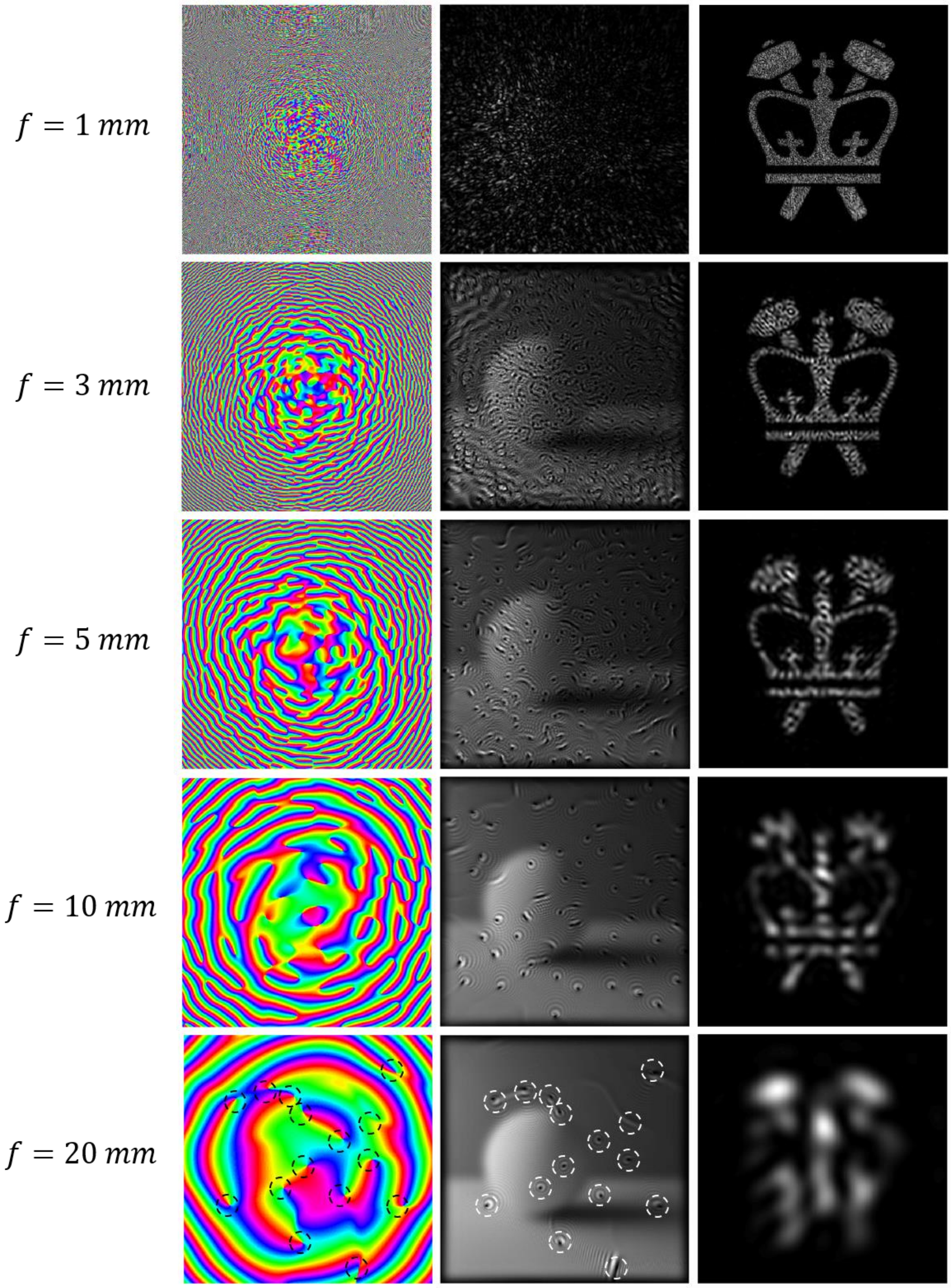

**Figure S12**. Comparison of the phase profiles (left column), simulated imaging of the metasurface plane (middle column), and simulated reconstructions of the object plane (right column) for various object plane distances, $f$, from the metasurface. Note trade-off in image quality at the two planes, increasing with $f$ for the metasurface plane and decreasing with $f$ for the object plane. Note too the correspondence between the artifacts at the metasurface plane with phase singularities in the phase profile (highlighted in the last row). This is understood by the destructive interference upon summation of pixels of approximately equal amplitudes, but varying across $2\pi$ in phase along the dashed contours shown circling the singularities.

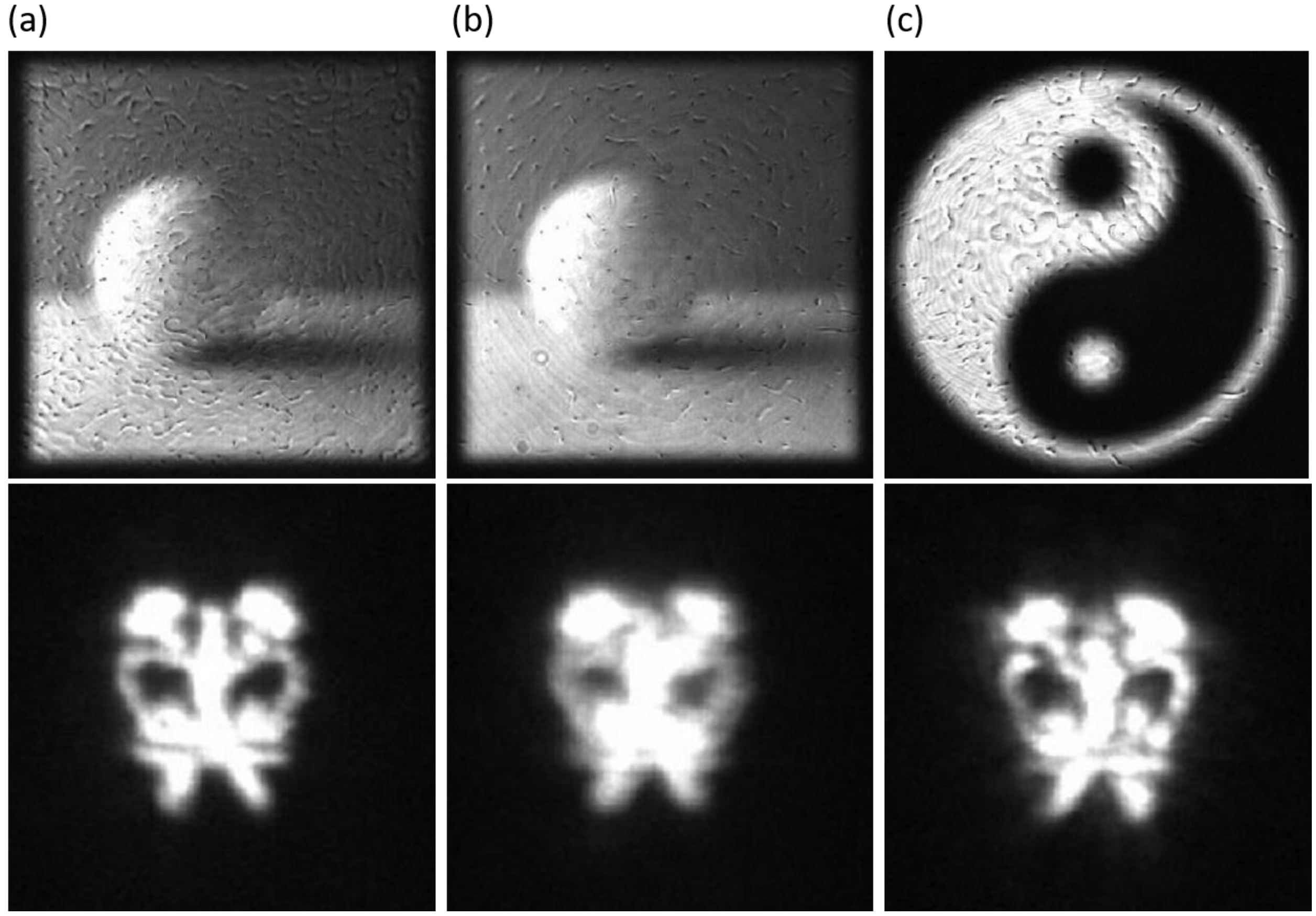

**Figure S13**. Experimental reconstructions of PA holograms using the modified GS algorithm, using LED illumination. (a) The device in Figure 6 of the main text reconstructed at the metasurface plane (top) and object plane (bottom), which is at $f = 3\ mm$. (b) A device similar to that in (a) but with object plane at $f = 5\ mm$, representing a different trade-off point from Figure S12. Note that the metasurface plane image looks improved at the expense of the object plane image. (c) Additional experimental hologram, showing a binary image at the metasurface plane (top) but the same image at the object plane (bottom), where $f = 3\ mm$.